\documentclass[aps,prd,twocolumn,groupedaddress,sort&compress]{revtex4-2}

\usepackage{natbib}
\usepackage{cancel}
\usepackage{pifont} 
\usepackage{amsthm}         
\usepackage{amsmath} 	   
\usepackage{amssymb}       
\usepackage{amsfonts}
\usepackage{graphicx} 	    
\usepackage{verbatim}
\usepackage{color}
\usepackage{colortbl}
\usepackage{float}
\usepackage{multirow}
\usepackage{booktabs}
\usepackage{makecell}

\definecolor{babyblue}{rgb}{0.54, 0.81, 0.94}
\definecolor{corn}{rgb}{0.98, 0.93, 0.36}

\begin{document}

\title{Thawing Quintessence: \\Priors, evidence, and likely trajectories}

\author{David Shlivko}
\email{dshlivko@princeton.edu}
\affiliation{Department of Physics, Princeton University, Princeton, NJ 08544, USA}

\date{\today}

\begin{abstract}
We perform a Bayesian comparison between thawing quintessence and a cosmological constant, incorporating theoretically motivated priors on the phenomenological Pad\'e-w parameters used to model thawing dynamics. We find that thawing quintessence is consistently preferred over a cosmological constant when combining BAO data from DESI DR2 and CMB data from Planck+ACT with any of the major supernova compilations, including the recently updated DES-Dovekie sample.
This preference is not sensitive to our choice of prior, but it is contingent on the inclusion of supernovae in the analysis.
We comment on the consistency between various information criteria and Bayesian evidence ratios, finding that the Deviance Information Criterion (DIC) tracks the Bayesian evidence more reliably than either the Akaike Information Criterion (AIC) or the Bayesian Information Criterion (BIC).
Finally, we use observational likelihoods to identify which thawing trajectories are compatible with the available data, independently of theoretical priors.
\end{abstract}

\maketitle

\section{Introduction}
Increasingly precise measurements of the cosmic expansion history \cite{SN:pantheon, SN:union, SN:DES_collab, DESI:2024vi, desi_cosmo_2025, popovic_dark_2025} have recently begun to show signs of a discrepancy with the standard cosmological model ($\Lambda$CDM). Follow-up analyses (\emph{e.g.,} \cite{lodha_extended_2025, shlivko_optimal_2025, alestas_desi_2025, cline_simple_2025, akrami_has_2025, de_souza_thawing_2025, payeur_observations_2025, park_updated_2025}) have found that these data can be better fit by thawing quintessence theories---arguably the simplest class of dynamical dark energy---in which the cosmological constant $\Lambda$ is replaced by a canonical scalar field whose equation of state increases from a past-asymptotic value of $w = -1$. The fit to the data is further improved in more elaborate theories that can produce or mimic a phantom-like regime ($w < -1$) at early times, including theories with non-canonical kinetic terms \cite{chen_quintessential_2025, goldstein_monodromic_2025, koutroulis_uv-complete_2026}, non-minimal gravitational couplings \cite{wolf_assessing_2025, wolf_cosmological_2025, brax_weinbergs_2025, adam_comparing_2025, lopez_non-minimally_2025, nojiri_phantom_2025, nojiri_apparent_2026}, non-standard dark matter evolution \cite{yang_probing_2025, kumar_evidence_2025, giani_matter_2025, chen_evolving_2025}, and interacting dark sectors \cite{chen_quintessential_2025, andriot_phantom_2025, khoury_apparent_2025, bedroya_evolving_2025, toomey_kinetic_2025, giani_novel_2025}. 
In a Bayesian framework, these improvements in fit must be considered together with the prior probabilities assigned to each theory, as well as the prior probability densities assigned to each theory's parameters, in order to assess which description of dark energy is most likely \emph{a posteriori}. The aim of this work is to perform such a prior-informed Bayesian comparison between $\Lambda$CDM and the class of thawing quintessence theories as a whole, setting the stage for future comparisons to some of the more elaborate dark energy theories mentioned above. 

When an entire class of theories is parameterized by an abstract phenomenological model, it can be especially difficult to quantify the prior credences that should be assigned to its parameter values. Despite being common practice, it is precarious to assume that these priors are uniform without justifying why a \emph{particular} set of parameters, and not those of some re-parameterization of the same class of theories, should be uniformly distributed. Fortunately, the Pad\'e-w parameterization of thawing quintessence \cite{shlivko_optimal_2025} can overcome this issue by virtue of having physically meaningful parameters and extraordinary precision in modeling the dynamics of microphysical quintessence theories (see Refs. \cite{wolf_underdetermination_2023, wolf_scant_2024, wolf_assessing_2025} for an alternative approach using a quadratic approximation to quintessence potentials). In Sec. \ref{s_pade}, we will show how a theoretically motivated prior on the Pad\'e-w parameter space can be constructed based on the attractor dynamics that govern thawing quintessence \cite{cahn_field_2008} and the UV-consistency constraints imposed by the refined de Sitter conjecture \cite{Ooguri:2018wrx, garg_bounds_2019, agrawal_dark_2019}. 

Equipped with this prior, we use the data and methodologies outlined in Sec. \ref{s_data} to compute the Bayesian evidence ratio between $\Lambda$CDM and thawing quintessence. For comparison, we perform the same computation using uniform priors on the Pad\'e-w parameter space, and we find that the results do not differ significantly given the current level of precision in measurements. To help inform future analyses, we compare these Bayesian evidence ratios to differences in various information criteria between $\Lambda$CDM and thawing quintessence. Information criteria are simple and computationally inexpensive measures of a model's performance that balance its maximum likelihood against its number of parameters (or effective degrees of freedom). Our results show that the number of effective degrees of freedom attributable to the two Pad\'e-w parameters depends strongly on the dataset used to constrain them, and we will see that the Deviance Information Criterion---which accounts for this data-dependence---can reliably predict which model will have the greater Bayesian evidence. All of these model-comparison results are presented in Sec. \ref{s_MC}.

In Sec. \ref{s_PE}, we focus on observational constraints within the class of thawing quintessence theories, and we reconstruct the most likely trajectories of the dark energy equation of state based solely on fit to observational data (disregarding theoretical priors). We also present joint posterior distributions that illustrate the dependence of the Hubble constant $H_0$ and matter fraction $\Omega_m$ on the Pad\'e-w parameters. We conclude with a summary and discussion of our results 
in Sec. \ref{s_conclusions}.

\section{Thawing quintessence priors}\label{s_pade}

We will model the dynamics of thawing quintessence theories using the Pad\'e-w parameterization from Ref. \cite{shlivko_optimal_2025}:
\begin{equation}\label{e_pade}
	\epsilon_\text{pad\'e}(z) = \frac{3\epsilon_0}{3 + \eta_0 (z^3+3z^2+3z)}.
\end{equation} 
Here, the equation of state $\epsilon = \frac{3}{2}(1+w) = \frac{3}{2}(1+P/\rho)$ models the ratio between the quintessence field's pressure $P$ and its energy density $\rho$ as a function of redshift $z$. This model has a one-dimensional $\Lambda$CDM limit corresponding to $\epsilon_0 = 0$. Despite originating from a phenomenological Pad\'e expansion of $w$ \cite{PADE_alho}, this version of the parameterization is expressed in terms of two physically meaningful parameters, $\epsilon_0 = \epsilon_\text{pad\'e}(0)$ and $\eta_0 = d\ln\epsilon_\text{pad\'e}/dN|_{z=0}$, where we have defined $N \equiv \ln(a) = -\ln(1+z)$ as the logarithm of the scale factor $a$.

A key advantage of the Pad\'e-w model over alternative parameterizations is its reliability in precisely capturing the dynamics of a broad variety of microphysical thawing quintessence theories \cite{shlivko_optimal_2025}. That is, if dark energy is truly a thawing quintessence field with equation of state $\epsilon_\text{true}(z)$, one can trust that the best-fit $\epsilon_\text{pad\'e}(z)$ matches the true evolution to very good approximation. This property allows us to identify probabilities associated with the phenomenological parameters $\epsilon_0$ and $\eta_0$ with credences about the true value of the equation of state today and its true rate of change. For alternative thawing parameterizations that are not so robust, this identification would be less reliable, and for broader parameterizations such as the Chevallier–Polarski–Linder (CPL) form $w(a) = w_0 + w_a(1-a)$, the correspondence between the best-fit $\{w_0, w_a\}$ values and the true behavior of the equation of state today breaks down even more dramatically \cite{assessing, shlivko_optimal_2025}. The precision of the Pad\'e-w parameterization therefore makes it particularly suitable for both assigning meaningful priors and interpreting the resulting posteriors and evidences. 

The robustness of the Pad\'e-w parameterization has been validated across
a variety of thawing quintessence scenarios satisfying $\epsilon_0 \lesssim 3/2$ (equivalently, $w \leq 0$ today) and $\eta_0 \lesssim 100$. In more extreme scenarios, however, the accuracy of the Pad\'e-w parameterization may falter \cite{shlivko_optimal_2025}. As a result, we will restrict attention in this work to the modest regime of thawing quintessence obeying the bounds above. Within this domain, we can determine a prior probability density by applying our knowledge of general principles that guide the evolution of scalar fields and by appealing to Swampland conjectures that restrict the space of effective field theories with viable UV completions. When setting our priors, we will distinguish between two qualitatively different regimes in the Pad\'e-w parameter space: The smooth regime, $\eta_0 < 3$, in which $d\ln\epsilon_\text{pad\'e}/dN$ slowly decays from $3$ to $0$ over time, and the spiky regime, $\eta_0 > 3$, in which $d\ln\epsilon_\text{pad\'e}/dN$ diverges upwards from $3$ toward $+\infty$ in finite time.

We begin by assuming a canonical scalar field with equation of state
\begin{equation}
	\epsilon = \frac{3X}{X+V},
\end{equation}
where $X = \frac{1}{2}\dot{\varphi}^2$ is the field's kinetic energy density and $V(\varphi)$ is its potential.
When $\epsilon \ll 1$ (\emph{e.g.,} in the far past, when the field is frozen by Hubble friction), one can use the field's equation of motion to show that
\begin{equation}\label{epsp}
	\frac{d\ln\epsilon}{dN} = 3\left(\left|\frac{V_{,\varphi}}{V}\right|\sqrt{\frac{2\Omega_\varphi(N)}{\epsilon(N)}}-2\right),
\end{equation}
where $\Omega_\varphi \approx V/(3H^2)$ relates the energy density in the field to the Hubble parameter $H = dN/dt$. Note that dimensionful quantities are implicitly written in units of the reduced Planck mass $M_{pl} = \sqrt{\hbar c / (8\pi G)}$. As long as we are in the slow-roll quintessence regime, where $V$ and $V_{,\varphi} \equiv dV/d\varphi$ are effectively constant, we can write
\begin{equation}\label{epspp}
	\frac{d^2\ln\epsilon}{dN^2} 
	=	\frac{3}{2}\left|\frac{V_{,\varphi}}{V}\right| \sqrt{\frac{2\Omega_\varphi}{\epsilon}}\left((\ln\Omega_\varphi)'-(\ln\epsilon)'\right),
\end{equation}
with primes denoting derivatives with respect to $N$. During matter domination ($\Omega_\varphi \ll \Omega_m$), one has that
\begin{equation}
	\Omega_\varphi \approx \frac{V}{\rho_m} \propto a^3 = e^{3N},
\end{equation}
and so $(\ln\Omega_\varphi)' = 3$ is a constant. This result, in combination with Eq. (\ref{epspp}), implies that there exists an attractor fixed point for $(\ln\epsilon)'$, namely, when it is also equal to $3$. Therefore, for any initial value $0 < \epsilon \ll 1$ at the onset of matter domination (characteristic of thawing quintessence), one will generically expect the scalar field's equation of state to approach the following attractor trajectory during the matter-dominated phase:
\begin{equation}\label{evol}
	\ln\epsilon(N) = 3N + const. \text{\qquad [Matter domination]}.
\end{equation}
We note here that the Pad\'e-w parameterization has the beautiful property of automatically satisfying $d\ln\epsilon/dN \to 3$ in the limit $N \to -\infty$.

While the slope of $\ln\epsilon(N)$ during matter-domination is independent of the quintessence field's potential, the constant offset in Eq. (\ref{evol}) is not. Instead, by combining Eq. (\ref{epsp}) with Eq. (\ref{evol}), one can show that the specific evolution of $\epsilon(N)$ is governed by
\begin{equation}\label{e_epsofn}
	\epsilon(N) = \frac{2}{9}\left(\frac{V_{,\varphi}}{V}\right)^2\Omega_\varphi(N) \text{\qquad [Matter domination]}.
\end{equation}
Near the end of matter domination, as $\Omega_\varphi$ approaches $\mathcal{O}(1/2)$, we see that $\epsilon \sim 0.1(V_{,\varphi}/V)^2$. 

According to the refined de Sitter conjecture \cite{Ooguri:2018wrx, garg_bounds_2019, agrawal_dark_2019}, consistency with quantum-gravitational UV completion requires that either $V_{,\varphi}/V \gtrsim 1$ 
or $V_{,\varphi\varphi}/V \lesssim -1$. In the spiky regime of parameter space, where $\ln\epsilon$ changes rapidly over time, even values of $\epsilon_0 \sim \mathcal{O}(1)$ correspond to scenarios with $\epsilon \lesssim 0.1$ near the end of matter domination, indicating that $V_{,\varphi}/V \lesssim 1$ and requiring a concave potential with $V_{,\varphi\varphi}/V \lesssim -1$.

We can parameterize such a concave (hilltop) potential as 
\begin{equation}
	V(\varphi) \approx 2m^2f^2 - \frac{1}{2}m^2\varphi^2  \qquad [\eta_0 > 3],
\end{equation}
using notation that facilitates comparison to axion quintessence models. 
Let us also assume that $\varphi < \mathcal{O}(f)$, so that the potential today is of order $m^2f^2$. 
According to the refined de Sitter conjecture, we must have $f^2 \approx -V/V_{,\varphi\varphi} \lesssim 1$ in the spiky regime of parameter space. Under this condition, for as long as $\epsilon \lesssim 3$, the field's late-time equation of state (during dark-energy domination) obeys \cite{shlivko_tcc}
\begin{equation}
	\eta \equiv \frac{d\ln\epsilon}{dN} \approx \frac{2}{f} \text{\qquad [DE domination; $\eta_0 > 3$]}.
\end{equation}
(Note that in disfavored cases corresponding to $f \gtrsim 1$, one instead finds that $\eta_0 \approx f^{-2} \lesssim 1$, so these models are not relevant to our analysis of the spiky regime.) Treating $f$ as a scale parameter of the model that can take on values across many orders of magnitude (as in the string axiverse; see, \emph{e.g.,} Ref. \cite{mehta_superradiance_2021}), we can assign to it a Jeffreys prior scaling as $1/f$. The prior on $\eta_0$ will thus similarly obey
\begin{equation}\label{eta0prior_spiky}
	\pi(\eta_0) \propto \eta_0^{-1} \qquad [\eta_0 > 3].
\end{equation}
Moreover, given a fixed value of $f$, the equation of state in this scenario is set by \cite{shlivko_tcc}
\begin{equation}
	\epsilon \approx \frac{\varphi^2}{f^2} \text{\qquad [DE domination; $\eta_0 > 3$]}.
\end{equation}
Taking the initial field value to have been uniformly distributed, and allowing the field to have scaled up by a fixed multiplicative factor during the DE-dominated phase (where $\ln\epsilon$, and hence $\ln\varphi$, evolve linearly in $N$), we conclude that the present-day field value is also uniformly distributed. As a result, the present-day equation of state $\epsilon_0 \propto \varphi_0^2$ given a fixed value of $\eta_0$ is distributed as 
\begin{equation}
	\pi(\epsilon_0|\eta_0) \propto \epsilon_0^{-1/2} \qquad [\eta_0 > 3].
\end{equation}

Turning to the smooth regime with $\eta_0 < 3$, we begin by noting that 
 the majority of the prior mass within this range should be concentrated at values of $1/2 \lesssim \eta_0 < 3$. This is because smaller values of $\eta_0$ would predict the rate of change of the equation of state at early times,
\begin{equation}
	\frac{d\ln\epsilon_\text{pad\'e}}{dN} = \frac{\eta_0 a^{-3}}{1 + \eta_0(a^{-3}-1)/3},
\end{equation}
to fall far short of the expectation from Eq. (\ref{evol}) that $d\ln\epsilon/dN \approx 3$ throughout the entire matter-dominated era. In contrast, the range $1/2 \lesssim \eta_0 < 3$ corresponds to well-motivated exponential (and other locally convex) potentials \cite{shlivko_optimal_2025}, whose prior mass should be comparable in magnitude to that of the spiky regime. A simple uniform prior on $\eta_0$ in this regime suffices to approximately match these expectations:
\begin{equation}\label{eta0prior_smooth}
	\pi(\eta_0) \propto 1 \qquad [\eta_0 \leq 3].
\end{equation}
For the prior on $\epsilon_0$, we continue to use
\begin{equation}
	\pi(\epsilon_0|\eta_0) \propto \epsilon_0^{-1/2} \qquad [\eta_0 \leq 3]
\end{equation}
so that the prior density function is continuous across the smooth-spiky boundary. Referring back to Eq. (\ref{e_epsofn}), the physical implication of this choice of prior is that the expected distribution of $|V_{,\varphi}/V|$ between $0$ and ${O}(3)$ is roughly uniform near the end of matter domination in smooth-thawing scenarios where $V_{,\varphi\varphi}/V$ tends to be positive. In true Bayesian fashion, the corner of parameter space with $|V_{,\varphi}/V| \lesssim 1$ and $V_{,\varphi\varphi}/V > 0$, which is nominally disallowed by the refined de Sitter conjecture, ends up with a nonzero prior density that comprises a small ($< 10\%$) fraction of the total prior mass. 

In sum, we arrive at the following priors on the Pad\'e-w parameter space:
\begin{align} \label{e_prioreps}
  \pi(\epsilon_0) &\propto \; \epsilon_0^{-1/2}, \quad 0 \le \epsilon_0 \le \tfrac{3}{2},\\[6pt] \label{e_prioreta}
  \pi(\eta_0)     &\propto \;
    \begin{cases}
      1,          & 0 \le \eta_0 \le 3,\\[4pt]
      3\eta_0^{-1},& 3 < \eta_0 \le 100.
    \end{cases}
\end{align}
We emphasize that these priors are not meant to represent an exact, objective truth. Rather, they are an approximate quantification of certain beliefs about quintessence fields' dynamics and their UV consistency, and readers are welcome to adjust these priors using additional or alternative beliefs. The key point is that, in contrast to a uniform prior, we have clarified which beliefs this particular prior encapsulates, and we are not \emph{unintentionally} introducing artificial preferences for certain regions of the parameter space.

\section{Data and methodology}\label{s_data}

We perform statistical comparisons between $\Lambda$CDM and thawing quintessence using likelihoods from the following datasets: 

\begin{enumerate}
	\item \textbf{CMB.} 
	We use the \texttt{P-ACT} combination of power spectra from the Planck satellite and ACT, which includes low-$\ell$ TT data from the Planck PR3 likelihood \cite{planck_2020_like}, low-$\ell$ EE data from the Sroll2 likelihood \cite{pagano_reionization_2020}, partial high-$\ell$ data from Planck PR3 restricted to $\ell < 1000$ in TT and $\ell < 600$ in TE/EE, and the ACT DR6 dataset \cite{naess_atacama_2025, louis_atacama_2025}.
	We also include a combination of CMB lensing data from ACT \cite{ACT_lensing1, ACT_lensing2, ACT_lensing3} and Planck's PR4 (NPIPE) maps \cite{carron_Planck_lensing}, choosing the \texttt{actplanck\_baseline} likelihood variant.
	\item \textbf{BAO.} We use the DESI DR2 observations of galaxies, quasars, and the Lyman-alpha forest \cite{desi_forest_2025, desi_cosmo_2025, andrade_validation_2025, casas_validation_2025, brodzeller_construction_2025}, measuring the angular diameter distance $D_M(z)$, the Hubble distance $D_H(z)$, or the angle-averaged quantity $D_V(z) \equiv (zD_M(z)^2D_H(z))^{1/3}$ in seven redshift bins. The individual DESI observations span $0.1 \leq z \leq 4.2$, while the binned effective redshifts range from $z = 0.295$ to $z = 2.330$. 
	\item \textbf{Supernovae.} We compare results for combinations of CMB and BAO data with various supernova catalogs from PantheonPlus, Union3, and DES:
	\begin{itemize}
		\item The PantheonPlus compilation \cite{SN:pantheon} contains 1701 measurements of 1550  Type Ia supernovae from redshifts $0.001 < z < 2.26$. (Note that the 111 measurements at $z \leq 0.01$ are filtered out in statistical analyses.) 
		\item The Union3 compilation \cite{SN:union} groups 2087 Type Ia supernova measurements from redshifts $0.01 < z < 2.26$ into twenty-two effective redshift bins ranging from $z = 0.05$ to $z = 2.26$. 
		\item The DES-SN5YR compilation \cite{sanchez_dark_2024, vincenzi_dark_2024, SN:DES} combines 1635 photometrically classified Type Ia supernovae from redshifts $0.1 < z < 1.3$ with 194 spectroscopically confirmed lower-redshift supernovae ($0.025 < z < 0.1$) from older catalogs. 
		\item The DES-Dovekie compilation \cite{popovic_dark_2025} is a re-calibrated and updated version of the DES-SN5YR compilation, which now comprises 1623 high-redshift and 197 low-redshift Type Ia supernovae.
	\end{itemize}
\end{enumerate}

\begin{table}[t]
\centering
	\begin{tabular}{|l|l|}
		\hline
		Parameter & Prior\\
		\hline
		$\ln(10^{10}A_s)$ & $\mathcal{U}[2.9, 3.2]$ \\
		$n_s$ & $\mathcal{U}[0.9, 1.05]$\\
		$\tau$ & $\mathcal{U}[0.02, 0.1]$\\
		$100\theta_\text{MC}$ & $\mathcal{U}[1.03, 1.05]$\\
		$\omega_b$ & $\mathcal{U}[0.02, 0.025]$\\
		$\omega_c$ & $\mathcal{U}[0.1, 0.14]$\\
		$A_\text{ACT}$ & $\mathcal{N}[1.0, 0.003]$\\
		$P_\text{ACT}$ & $\mathcal{U}[0.9, 1.1]$\\
		$\epsilon_0$ & Eq. (\ref{e_prioreps}) or \,$\mathcal{U}[0, 1.5]$\\
		$\eta_0$ & Eq. (\ref{e_prioreta}) or \,$\mathcal{U}[0, 100]$\\
		\hline
	\end{tabular}
	\caption{Parameters used in our statistical analyses and their priors. Uniform and normal distributions are denoted respectively by $\mathcal{U}[a, b]$ and $\mathcal{N}[\mu, \sigma]$. $A_s$ is the amplitude of the primordial power spectrum, $n_s$ is the spectral index, $\tau$ is the optical depth, and $\theta_\text{MC}$ is the angular size of the sound horizon at recombination. The physical densities of baryons (b) and cold dark matter (c) are defined as $\omega_i \equiv \Omega_i h^2$, with $h \equiv H_0 / (100\text{ km/s/Mpc})$. The variables $A_\text{ACT}$ and $P_\text{ACT}$ correspond to ACT's dipole and polarization calibration parameters. Finally, the dark energy equation of state in thawing quintessence models is defined by the Pad\'e-w parameters $\{\epsilon_0, \eta_0\}$ per Eq. (\ref{e_pade}).}\label{t_priors}
\end{table}

We perform all of our statistical analyses within \texttt{cobaya} \cite{cobaya1, cobaya2}, making use of the \texttt{CAMB} Boltzmann solver \cite{camb1, camb2}. Within \texttt{CAMB}, we use the parameterized post-Friedmann approach \cite{fang_ppf} to compute dark energy perturbations, and we modify the 2016 \texttt{HMcode} algorithm \cite{mead_hmcode_2015, mead_accurate_2016} for computing non-linear matter power spectra to use the (tabulated) Pad\'e-w parameterization input into CAMB instead of the default CPL parameterization.
For accurate estimates of maximum likelihood, we run multiple parallel iterations of the \texttt{iminuit} maximizer \cite{iminuit}. To compute posterior-weighted averages of the log-likelihood (or related quantities), we employ the Metropolis-Hastings MCMC sampler with dragging \cite{metropolis1, metropolis2, metropolis_drag}. The convergence of MCMC chains is determined using the Gelman-Rubin statistic \cite{gelman_inference_1992}, for which we require $R - 1 < 0.01$, and plots of the MCMC results are constructed using \texttt{GetDist} \cite{getdist}. For accurate estimates of the Bayesian evidence, we use the Polychord sampler \cite{handley_polychord_2015, handley_polychord_2015-1}, with a \texttt{precision\_criterion} of $0.001$, \texttt{nprior = 20nlive}, and \texttt{nlive = 40d}.

The priors for variable parameters in our analyses are shown in Table \ref{t_priors}. For computational efficiency with Polychord, we choose relatively narrow priors for cosmological parameters that are well-constrained independently of the nature of dark energy; we ensure that the posteriors for these parameters ($A_s$, $n_s$, $\tau$, $\theta_\mathrm{MC}$, $\omega_b$, and $\omega_c$) all still lie well within their respective priors. We compute Bayesian evidence ratios using both informed priors and uniform priors on the Pad\'e-w parameters $\{\epsilon_0, \eta_0\}$ for comparison. Throughout this work, we assume a spatially flat universe, a single massive neutrino with $m_\nu = 0.06$ eV, and $N_\text{eff} = 3.044$ total neutrino species. 

\begin{table*}[t]
    \centering
    \label{tab:example}
    \begin{tabular}{*{9}{c}}
        \toprule
        \text{Dataset} & $\Delta\ln Z$   & $\Delta$AIC & $\Delta$BIC & $\Delta$$\widehat{\text{DIC}}$ & $\Delta\widetilde{\text{DIC}}$ & $\Delta\ln L_\text{max}$ & $\Delta\widehat{d}$ & $\Delta\widetilde{d}$ \\
        \midrule
        \text{DESI+CMB} & -0.7 (-0.7) & 3.8 & 13.3 & 0.8 (0.6) & 0.5 (1.1) & 0.1 & 0.5 (0.4) & 0.4 (0.7) \\
        \text{DESI+CMB+PP} & 0.5 (0.8) & 0.1 & 11.7 & -0.4 (-1.4) & -1.7 (-1.8) & 1.9 & 1.7 (1.2) & 1.1 (1.0) \\
        \text{DESI+CMB+U3} & 1.7 (2.1) & -3.2 & 6.3 & -2.3 (-3.3) & -3.7 (-5.7) & 3.6 & 2.5 (2.0) & 1.8 (0.8) \\
        \text{DESI+CMB+D5$^*$} & 5.0 (5.5) & -9.9 & 1.9 & -9.3 (-10.6) & -9.1 (-11.3) & 6.9 & 2.3 (1.7) & 2.4 (1.3) \\
        \text{DESI+CMB+DD} & 1.5 (2.4) & -3.1 & 8.7& -2.5 (-4.1) & -3.4 (-4.9) & 3.5 & 2.3 (1.5) & 1.8 (1.1) \\
        \bottomrule
    \end{tabular}
     \caption{\label{t_results}Statistical results comparing thawing quintessence to $\Lambda$CDM. Parentheses indicate results obtained using a uniform prior on the Pad\'e-w parameters, rather than the informed prior from Eqs. (\ref{e_prioreps}-\ref{e_prioreta}). A Bayesian preference for thawing quintessence corresponds to positive $\Delta \ln Z$, while an information-criterion-based preference for thawing quintessence corresponds to negative $\Delta IC$. Also shown are changes in maximum likelihood (always positive semi-definite, since $\Lambda$CDM is a limit of thawing quintessence) and in the Bayesian model complexity $\widehat{d}$ and dimensionality $\widetilde{d}$, which count the effective degrees of freedom introduced by Pad\'e-w. Rows correspond to different data sets, with supernova samples abbreviated as PP (PantheonPlus), U3 (Union3), D5 (DES-SN5YR), and DD (DES-Dovekie). The uncertainties in stated values are estimated as either $\pm 0.1$ ($\Delta\ln L_\text{max}$, $\Delta\widehat{d}$, $\Delta\widetilde{d}$), $\pm 0.2$ ($\Delta$AIC, $\Delta$BIC), or $\pm 0.4$ ($\Delta\ln Z$, $\Delta$$\widehat{\text{DIC}}$, $\Delta\widetilde{\text{DIC}}$). \newline $^*$The legacy D5 supernova sample has recently been re-analyzed and superseded by DD. Results are shown here for both samples in the interest of comparison.}
\end{table*}

\section{Model comparison}\label{s_MC}
The main result of this work is a calculation of Bayesian evidence ratios between thawing quintessence and $\Lambda$CDM, which we show in the first column of Table \ref{t_results} for various combinations of datasets. The evidence of a model $M$ is defined as its prior-weighted average likelihood,
\begin{equation}
	Z_M \equiv \int d^n\theta P(\text{data} \mid M, \theta ) \cdot \pi(\theta\mid M),
\end{equation}
where $\theta$ represents the model's $n$-dimensional parameter space and $\pi(\theta\mid M$) is its joint prior probability density. In Table \ref{t_results}, we report the logarithm of the ratio $Z_\text{TQ}/Z_\Lambda$, which represents the Bayesian updating factor that multiplies a relative prior $\Pi_\text{TQ}/\Pi_\Lambda$ on the overall models to obtain the relative data-informed posteriors:
\begin{equation}
	\frac{P(\text{TQ}\mid \text{data})}{P(\Lambda\mid \text{data})} = \frac{\Pi_\text{TQ}}{\Pi_\Lambda} \cdot \frac{Z_\text{TQ}}{Z_\Lambda}.
\end{equation}
In other words, $\Delta\ln Z > 0$ means that one's credence for thawing quintessence relative to $\Lambda$CDM has increased in light of the data, and vice versa. We remind the reader that in our analysis, ``TQ'' refers specifically to the non-extreme subset of thawing quintessence theories satisfying $\epsilon_0 \leq 3/2$ and $\eta_0 \leq 100$. 

We see from Table \ref{t_results} that DESI+CMB data alone slightly \emph{decrease} the relative credence for thawing quintessence, while combinations of data that include the latest supernova samples \emph{increase} the relative credence by up to a factor of $\mathcal{O}(10)$. The legacy DES-SN5YR sample produced a stronger evidence ratio in excess of $\mathcal{O}(100)$, but the DES-Dovekie reanalysis brought the results more in line with those using the Union3 sample. Note that the evidence ratios computed using a uniform prior, shown in parentheses in Table \ref{t_results}, are qualitatively similar to those computed using the informed priors derived in Sec. \ref{s_pade}. 

The next four columns of Table \ref{t_results} show the differences in various information criteria between thawing quintessence and $\Lambda$CDM. These information criteria have been used to guide model selection \cite{liddle_information_2007} by quantifying the tradeoff between a model's improvement in fit and its added complexity (\emph{i.e.,} by quantifying the Occam penalty for models with more degrees of freedom). For example, the Akaike Information Criterion (AIC) is given by
\begin{equation}
	\text{AIC} = 2p - 2\ln L_\text{max},
\end{equation}
where $p$ is the number of parameters in a model $M$ and $L_\text{max} = \max_\theta\left[P(x\mid M, \theta)\right]$ is its maximum likelihood \cite{akaike_new_1974}. In extending the $\Lambda$CDM model to Pad\'e-w, we take $\Delta p = 2$, and differences in $\ln L_\text{max}$ are reported in Table \ref{t_results}. Because the AIC is a measure of information \emph{lost} by a given model, negative values of $\Delta$AIC indicate a preference for thawing quintessence. In particular, thawing quintessence is more likely than $\Lambda$CDM to minimize information loss by a factor of $\exp(-\Delta\text{AIC}/2)$. We see that the AIC-based model preferences are mostly aligned with the Bayesian evidence results, though they more heavily favor $\Lambda$CDM in the case with just DESI+CMB data, and they fail to capture the slight preference for thawing quintessence when adding in PantheonPlus supernovae.

The Bayesian Information Criterion 
\begin{equation}
	\text{BIC} = \ln(N)p - 2\ln L_\text{max}
\end{equation}
applies a stronger Occam penalty to the number of parameters $p$ based on the number of data points $N$ used to test the model. (Note that we use $N_\text{CMB} \approx 824$ as an effective measure of CMB data points, based on the binning strategies described in Ref. \cite{louis_atacama_2025} and consistency with $\chi^2$ values.) Despite being designed to approximate a model's marginal likelihood \cite{BIC_schwarz}, we see that the BIC's increased Occam penalty is far too severe, with even the clearest cases of evidence favoring thawing quintessence being rejected by the BIC in favor of $\Lambda$CDM. This is not surprising: The BIC approximation assumes that models are evaluated using $N$ independent and identically distributed data points and that priors on the models' parameters are ``unit-information'' multivariate Gaussians centered at the maximum-likelihood point. Neither of these assumptions are applicable to the present analysis, and the effective number of independent data points with the strongest constraining power on $\epsilon_0$ and $\eta_0$ is likely significantly smaller than the total number of data points $N$.

Finally, we define two versions of the Deviance Information Criterion \cite{spiegelhalter_bayesian_2002, spiegelhalter_deviance_2014},
\begin{equation}
	\widehat{DIC} = 2\widehat{d} - 2\ln L_\text{max}
\end{equation}
and
\begin{equation}
		\widetilde{DIC} = 2\widetilde{d} - 2\ln L_\text{max},
\end{equation}
which modify the AIC by replacing the parameter count $p$ with a Bayesian measure of the model's effective number of  degrees of freedom. To this end, one can either use the standard notion of Bayesian model complexity,
\begin{equation}
	\widehat{d} = 2\ln L_\text{max} - 2\langle \ln L \rangle,
\end{equation}
or the more recently introduced Bayesian model dimensionality,
\begin{equation}
	\widetilde{d} = \text{Var}\left[ \ln L \right],
\end{equation}
which has been found to be more accurate in the context of cosmological parameter estimation \cite{handley_quantifying_2019}.
In these expressions, averages and variances are weighted by posterior probabilities, making them straightforward to compute from MCMC chains. The increases in $\widehat{d}$ and $\widetilde{d}$ when extending $\Lambda$CDM to the Pad\'e-w parameterization are shown in the final two columns of Table \ref{t_results}. These increases are only of order $\Delta p = 2$ when $\epsilon_0$ and $\eta_0$ are well-constrained by the data; weaker constraints (as when using PantheonPlus supernovae or no supernova data at all) imply correspondingly fewer effective degrees of freedom. This dynamic allows both versions of the DIC to closely track the evidence ratios between thawing quintessence and $\Lambda$CDM, preferring the same model in all cases considered. We do notice that in some cases, the two definitions of the DIC may disagree on the quantitative extent of this preference, and $\Delta$DIC may be more sensitive to the choice of prior than $\Delta\ln Z$, underscoring the role of information criteria as qualitative indicators without the quantitative robustness of Bayesian evidence.

\begin{figure*}[p]
\begin{center}
	\includegraphics[width=0.9\textwidth]{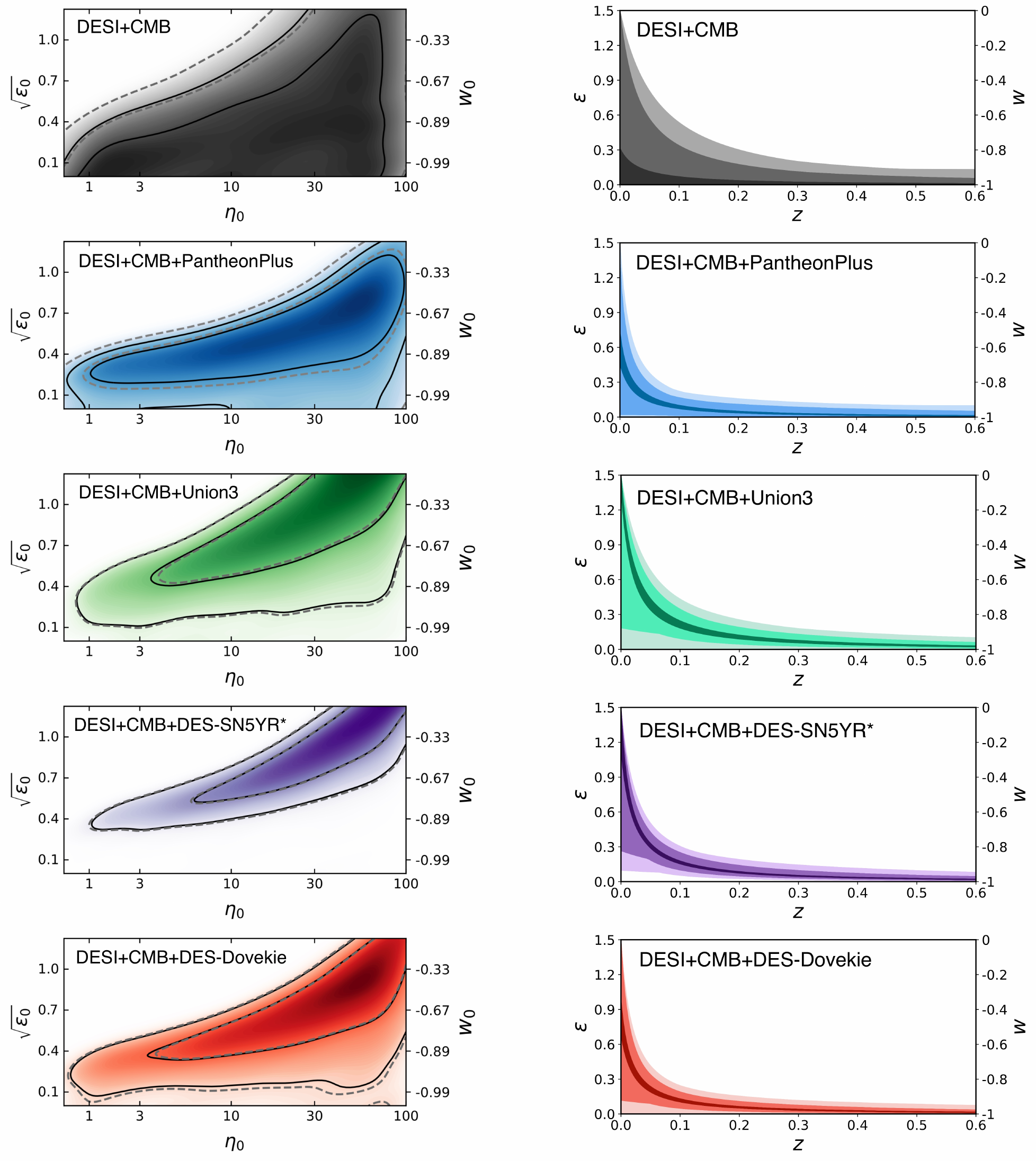}\vspace{-3mm}
	\caption{\emph{Left: }MCMC posterior densities for the Pad\'e-w parameters $\epsilon_0$ and $\eta_0$ according to combinations of DESI and CMB data with and without supernova samples. The linear scaling in $\sqrt{\epsilon_0}$ and the nonlinear scaling in $\eta_0$ are adopted so that the informed priors from Eqs. (\ref{e_prioreps}-\ref{e_prioreta}) appear uniform on the parameter space, allowing posteriors to directly represent marginal observational likelihoods. 
	We include a twin axis with (nonlinear) values of $w_0 = \frac{2}{3}\epsilon_0-1$ for ease of interpretation. Contours drawn in solid black contain 68\% and 95\% of the posterior mass, while contours drawn in dashed gray identify level sets with 32\% and 5\% of the maximum posterior density. Note that the $\Lambda$CDM limit of this parameterization corresponds to the one-dimensional boundary $\epsilon_0 = 0$.
	\newline \emph{Right: }High-likelihood evolutions of the dark energy equation of state $\epsilon(z)$ or $w(z) = \frac{2}{3}\epsilon(z)-1$ according to combinations of DESI and CMB data with and without supernova samples. From darkest to lightest, the shaded regions are reconstructed from $(\epsilon_0, \eta_0)$ combinations with $\geq 90\%$, $\geq 32\%$, and $\geq 5\%$ of the maximum marginal likelihood. The boundaries of the 32\% and 5\% regions correspond to the dashed gray contours on the Pad\'e-w parameter space shown in the corresponding panels on the left.
	\newline $^*$The DES-SN5YR supernova sample has recently been re-analyzed and superseded by DES-Dovekie. Results are shown for both samples in the interest of comparison.}\label{f_posteriors}
\end{center}
\end{figure*}

\section{Constraints on thawing dynamics}\label{s_PE}

Within the class of thawing dark energy theories, some dynamics are more likely than others---both from a theoretical perspective, as reflected in our priors, and due to observational constraints, as encoded in the likelihood function. The posterior probability densities computed via MCMC or nested sampling capture both of these effects, and they can be directly used to compare the integrated posterior masses of different regions of parameter space corresponding to different families of theories. 

In practice, however, one may wish to spot-check whether a specific theory (corresponding to a single point in parameter space) or a one-parameter family of theories (\emph{e.g.,} exponential potentials with varying steepness) is compatible with observational data. This is not a question about posterior masses (which are, in these cases, infinitesimal), but rather about observational likelihoods. 
To visualize the likelihoods corresponding to different points in the Pad\'e-w parameter space, we can simply plot the MCMC posterior distributions using rescaled axes along which the priors from Eqs. (\ref{e_prioreps}-\ref{e_prioreta}) appear uniform. In particular, our new axes will scale linearly in $\sqrt{\epsilon_0}$ and in the (piecewise) cumulative density function of $\eta_0$ (see left panels of Fig. \ref{f_posteriors}). We choose to visualize the likelihoods using rescaled axes rather than uniform priors on the original axes (as was done in Ref. \cite{shlivko_optimal_2025}) because the posterior densities are more nicely behaved and less sensitive to smoothing artifacts.

To directly identify which thawing dark energy theories are most compatible with the data, we have drawn gray, dashed contours in the left panels of Fig. \ref{f_posteriors} corresponding to 32\% and 5\% of the maximum posterior density. We reiterate that in these rescaled axes, posterior densities simply represent observational likelihoods marginalized over the remaining cosmological parameters. For perfectly Gaussian posteriors, these contours would reproduce the standard $1\sigma$ and $2\sigma$ regions containing 68\% and 95\% of the posterior mass (plotted for comparison as black, solid contours). For non-Gaussian distributions, however, these posterior-mass-based contours may fail to include points in the parameter space with appreciable likelihoods. This happens most noticeably for the DESI+CMB combination of data (top-left panel), where many points with more than $5\%$ of the maximum posterior density, for example, lie outside of the $95\%$-posterior-mass contour. For analyses including supernova data, the falloff in posterior density is sufficiently close to Gaussian for this distinction to be less significant. 

In the right panels of Fig. \ref{f_posteriors}, we use 90\%, 32\%, and 5\% max-marginal-likelihood contours to reconstruct the most likely trajectories of the dark energy equation of state in thawing-quintessence scenarios. These trajectories offer an alternative, visual approach for comparing predictions from (non-extreme) thawing quintessence theories to observational constraints. Theories that are compatible with these constraints can later be evaluated in more detail using a Bayesian analysis that accounts for their respective microphysical priors. 

\begin{figure*}[t!]
	\includegraphics[width=\textwidth]{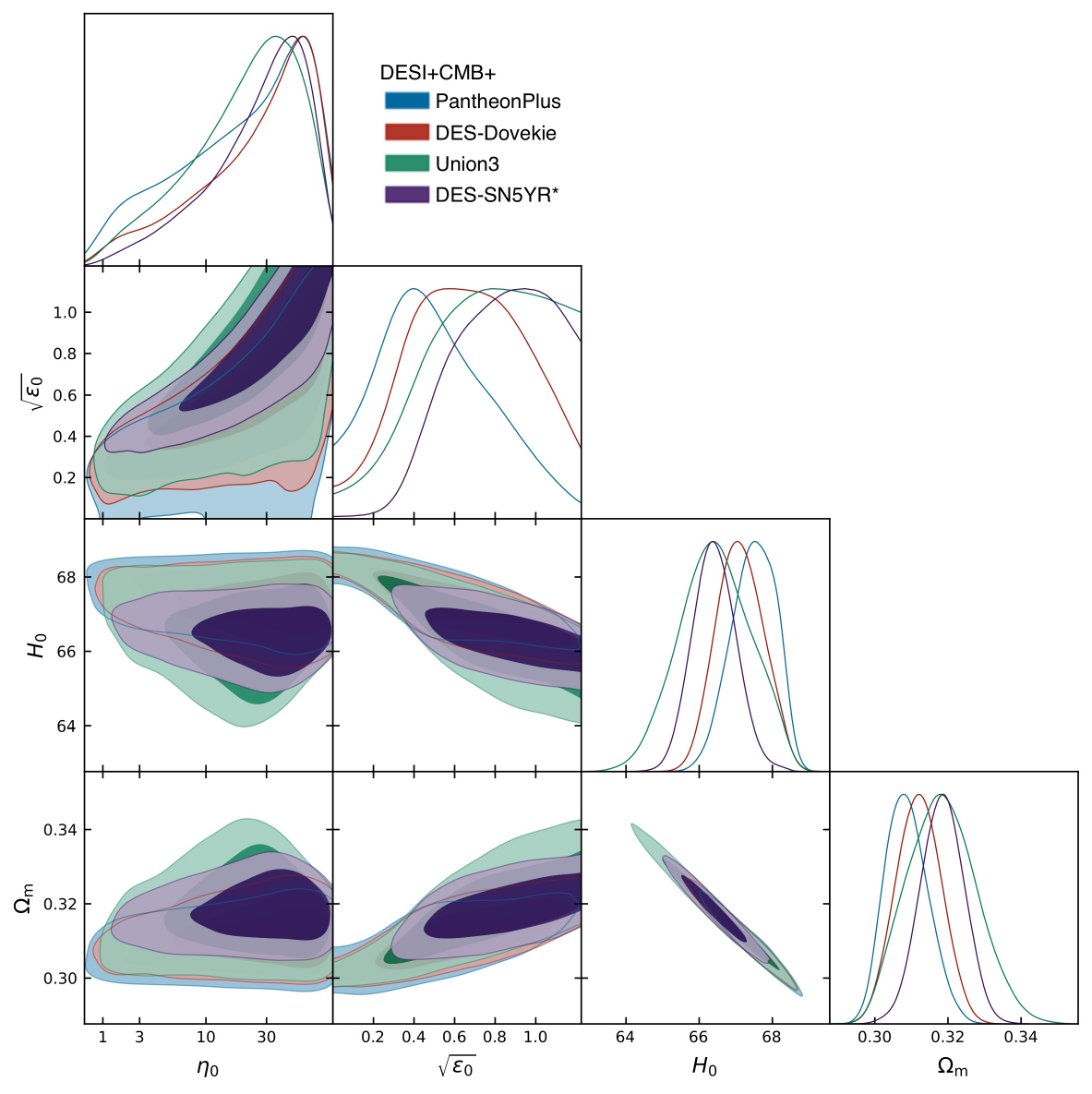}
	\caption{MCMC results showing marginalized posteriors and joint 68\% and 95\% credible regions for the Pad\'e-w parameters $\{\epsilon_0, \eta_0\}$, the Hubble constant $H_0$ (in km/s/Mpc), and the matter fraction $\Omega_m$. The linear scaling in $\sqrt{\epsilon_0}$ and the nonlinear scaling in $\eta_0$ are adopted so that the informed priors from Eqs. (\ref{e_prioreps}-\ref{e_prioreta}) appear uniform on the Pad\'e-w parameter space, as explained in the text. \newline $^*$The DES-SN5YR supernova sample has recently been re-analyzed and superseded by DES-Dovekie. Results are shown for both samples in the interest of comparison.}\label{f_cosmoparams}
\end{figure*}

While microphysical predictions for $\epsilon(z)$ depend on the present-day matter fraction $\Omega_m$, the MCMC results shown in Fig. \ref{f_cosmoparams} reveal that $\Omega_m$ is constrained to within $\sim 10\%$ even for large deviations from $\Lambda$CDM. Within this range, the predictions for $\epsilon(z)$ are expected to remain relatively robust, though one should always double-check that the assumed value of $\Omega_m$ is compatible with the resulting predictions for $\epsilon_0$ and $\eta_0$. The MCMC results also confirm that deviations from the $\Lambda$CDM limit ($\epsilon_0 = 0$) predict lower values of $H_0$ paired with higher $\Omega_m$, as expected from the kinematics of a diluting dark energy density. This effect exacerbates the tension between CMB-based constraints and late-time measurements of $H_0$ \cite{riess_comprehensive_2022, collaboration_local_2025, freedman_status_2025, banerjee_hubble_2021}.

\section{Discussion}\label{s_conclusions}

We began this work by constructing physically motivated priors on the Pad\'e-w parameter space, based on the attractor dynamics of thawing quintessence fields and UV-consistency constraints imposed by the refined de Sitter conjecture.
Using these priors, we computed the Bayesian evidence ratios between $\Lambda$CDM and the class of non-extreme thawing quintessence theories based on CMB and BAO data, as well as combinations of these data with various supernova samples. 
In the absence of supernovae, thawing quintessence offers almost no improvement in maximum likelihood over $\Lambda$CDM, resulting in a lower overall Bayesian evidence. However, tensions between BAO and supernova measurements (which, in $\Lambda$CDM, predict noticeably higher $\Omega_m$ \cite{desi_cosmo_2025}) \emph{can} be relieved in thawing-quintessence scenarios. This causes the evidence for thawing quintessence to turn positive when any supernova sample is included in the analysis. We note that in light of the DES-Dovekie analysis replacing DES-SN5YR, the Bayes factors favoring thawing quintessence are limited to $\lesssim \mathcal{O}(10)$.

Comparing these results to the Bayes factors obtained using uniform priors, 
we see that the informed prior did not affect the results by more than a factor of $\mathcal{O}(2)$, despite assigning significantly less prior mass to the high-$\eta_0$ (spiky-thawing) regime and more prior mass to the low-$\epsilon_0$ ($\Lambda$CDM-like) regime. To some extent, these differences balance each other out, as  current data show the highest likelihoods occurring in the high-$\eta_0$, high-$\epsilon_0$ corner of parameter space. Additionally, current constraints on $\epsilon_0$ and $\eta_0$ span a large fraction of the parameter space (see Fig. \ref{f_posteriors}), so any priors that do not artificially favor the ultra-low-likelihood corner with high $\epsilon_0$ and low $\eta_0$ should be expected to yield relatively similar prior-weighted average likelihoods. However, as the precision of data continues to improve and constraints on these parameters tighten, different priors will lead to increasingly different evidence ratios, and justifying one's priors will become increasingly important for presenting reliable quantitative measures of Bayesian model preference. We note that Bayesian analysis is particularly valuable for assessing multi-parameter models of thawing dark energy, given the difficulty of applying frequentist likelihood ratio tests to systems with degenerate 1D $\Lambda$CDM limits and approximate degeneracies in the spiky thawing regime. 

To help inform future analyses, we compared these Bayesian evidence ratios to differences in various information criteria between $\Lambda$CDM and thawing quintessence, finding that the Deviance Information Criterion (DIC) tracks the Bayesian evidence more consistently than either the Akaike Information Criterion (AIC) or the Bayesian Information Criterion (BIC). The slight improvement in consistency when using the DIC over the AIC demonstrates the effectiveness of applying either the Bayesian model complexity or Bayesian model dimensionality as a measure of the number of effective degrees of freedom in a model. Meanwhile, the failure of the BIC to act as a proxy for Bayesian evidence highlights the violated assumptions of i.i.d. data points and unit-information Gaussian priors in cosmological analyses.

Zooming in on the thawing quintessence models themselves, we showed in Fig. \ref{f_posteriors} the observational constraints on the Pad\'e-w parameters $\{\epsilon_0, \eta_0\}$ and on the corresponding evolution of the dark energy equation of state $\epsilon(z)$. Because standard $1\sigma$ and $2\sigma$ contours of highest posterior density are not invariant under reparameterization, and because they may misleadingly exclude observationally compatible models if posteriors are non-Gaussian, we presented these results using contours of constant marginal likelihood (equal to a fixed fraction of the maximum marginal likelihood). These results can help to identify at a glance whether a given microphysical theory is compatible with the available data, either by mapping the theory onto the phenomenological parameter space \cite{de_putter_calibrating_2008, wolf_underdetermination_2023, wolf_scant_2024, assessing, shlivko_optimal_2025} or by calculating its predictions for $\epsilon(z)$. Observationally compatible theories can then be studied in more detail and compared against each other through formal Bayesian analyses that properly account for the theories' respective priors. 

In future work, it will be interesting to extend the present Bayesian comparison between $\Lambda$CDM and thawing quintessence to encompass more elaborate dark energy theories, including those that can lead to (apparent) phantom-crossing behavior in the equation of state. This can be done either by simulating these theories directly or by using their microphysical dynamics to define informed priors on a shared phenomenological parameter space \cite{wolf_assessing_2025, wolf_cosmological_2025, toomey_how_2025, toomey_kinetic_2025, marsh_quintessence_2014, garcia-garcia_theoretical_2020, traykova_theoretical_2021}. The phenomenological approach will only work, however, if the chosen parameterization can accurately model the cosmological observables predicted by these theories (such that errors in the model are smaller than observational uncertainties).
Given the increasing precision of observational data, a rigorous calculation of Bayesian evidence may therefore require either direct simulation of the microphysics or, if many theories' phenomenologies are sufficiently similar, a precise ``phantom-crossing'' parameterization analogous to the role played by Pad\'e-w for thawing quintessence. 
In either case, such Bayesian analyses will help to determine quantitatively whether the hints of phantom crossing in current data justify a heightened credence in more elaborate theories of dark energy that may require new fine-tunings.

 \vspace{0.1in}
\noindent
\begin{center}
\textbf{ACKNOWLEDGMENTS}
\end{center}
I am grateful to Nicolas Patino, Paul Steinhardt, and William Wolf for useful conversations and comments on the manuscript. This work is supported in part by the DOE grant number DEFG02-91ER40671 and by the Simons Foundation grant number 654561. The simulations performed in this work utilized computational resources managed and supported by Princeton Research Computing, a consortium of groups including the Princeton Institute for Computational Science and Engineering (PICSciE) and the Office of Information Technology's High Performance Computing Center and Visualization Laboratory at Princeton University.

\bibliographystyle{apsrev4-2.bst}
\bibliography{ThawingBayes.bib}

@article{de_souza_thawing_2025,
	title = {Thawing quintessence and transient cosmic acceleration in light of {DESI}},
	volume = {112},
	url = {https://link.aps.org/doi/10.1103/2tjq-dtbc},
	doi = {10.1103/2tjq-dtbc},
	number = {8},
	urldate = {2025-11-18},
	journal = {Physical Review D},
	author = {de Souza, Rayff and Rodrigues, Gabriel and Alcaniz, Jailson},
	month = oct,
	year = {2025},
	zoteronote = {Publisher: American Physical Society},
	pages = {083533},
}

@misc{akrami_has_2025,
	title = {Has {DESI} detected exponential quintessence?},
	url = {http://arxiv.org/abs/2504.04226},
	doi = {10.48550/arXiv.2504.04226},
	urldate = {2025-11-18},
	publisher = {arXiv},
	author = {Akrami, Yashar and Alestas, George and Nesseris, Savvas},
	month = apr,
	year = {2025},
	note = {arXiv:2504.04226 [astro-ph]},
}

@article{lodha_extended_2025,
	title = {Extended dark energy analysis using {DESI} {DR2} {BAO} measurements},
	volume = {112},
	url = {https://link.aps.org/doi/10.1103/w4c6-1r5j},
	doi = {10.1103/w4c6-1r5j},
	number = {8},
	urldate = {2025-11-18},
	journal = {Physical Review D},
	author = {Lodha, K. and others},
	collaboration = {DESI Collaboration},
	month = oct,
	year = {2025},
	pages = {083511},
}

@article{cline_simple_2025,
	title = {Simple quintessence models in light of {DESI}-{BAO} observations},
	volume = {112},
	url = {https://link.aps.org/doi/10.1103/8z2m-nbv6},
	doi = {10.1103/8z2m-nbv6},
	number = {6},
	urldate = {2025-11-18},
	journal = {Physical Review D},
	author = {Cline, James M. and Muralidharan, Varun},
	month = sep,
	year = {2025},
	zoteronote = {Publisher: American Physical Society},
	pages = {063539},
}

@misc{alestas_desi_2025,
	title = {{DESI} constraints on two-field quintessence with exponential potentials},
	url = {http://arxiv.org/abs/2510.21627},
	doi = {10.48550/arXiv.2510.21627},
	urldate = {2025-11-18},
	publisher = {arXiv},
	author = {Alestas, George and Caldarola, Marienza and Ocampo, Indira and Nesseris, Savvas and Tsujikawa, Shinji},
	month = oct,
	year = {2025},
	note = {arXiv:2510.21627 [astro-ph]},
}

@misc{goldstein_monodromic_2025,
	title = {Monodromic {Dark} {Energy} and {DESI}},
	url = {http://arxiv.org/abs/2507.16970},
	doi = {10.48550/arXiv.2507.16970},
	urldate = {2025-10-23},
	publisher = {arXiv},
	author = {Goldstein, Samuel and Celoria, Marco and Schmidt, Fabian},
	month = jul,
	year = {2025},
	note = {arXiv:2507.16970 [astro-ph]},
	anzoteronote = {Comment: 4+5 figures, 8+7 pages, comments welcome!},
}

@article{garg_bounds_2019,
	title = {Bounds on slow roll and the de {Sitter} {Swampland}},
	volume = {2019},
	issn = {1029-8479},
	url = {https://doi.org/10.1007/JHEP11(2019)075},
	doi = {10.1007/JHEP11(2019)075},
	language = {en},
	number = {11},
	urldate = {2025-11-18},
	journal = {Journal of High Energy Physics},
	author = {Garg, Sumit K. and Krishnan, Chethan},
	month = nov,
	year = {2019},
	pages = {75},
}

@article{agrawal_dark_2019,
	title = {Dark energy and the refined de sitter conjecture},
	volume = {2019},
	issn = {1029-8479},
	url = {https://doi.org/10.1007/JHEP06(2019)103},
	doi = {10.1007/JHEP06(2019)103},
	language = {en},
	number = {6},
	urldate = {2025-11-18},
	journal = {Journal of High Energy Physics},
	author = {Agrawal, Prateek and Obied, Georges},
	month = jun,
	year = {2019},
	pages = {103},
}

@article{wolf_cosmological_2025,
	title = {Cosmological constraints on {Galileon} dark energy with broken shift symmetry},
	volume = {113},
	url = {https://link.aps.org/doi/10.1103/bxvj-bsv1},
	doi = {10.1103/bxvj-bsv1},
	number = {2},
	urldate = {2026-04-05},
	journal = {Physical Review D},
	author = {Wolf, William J. and Ferreira, Pedro G. and García-García, Carlos},
	month = jan,
	year = {2026},
	note = {Publisher: American Physical Society},
	pages = {023551},
}

@misc{toomey_kinetic_2025,
	title = {Kinetic {Mixing} and the {Phantom} {Illusion}: {Axion}-{Dilaton} {Quintessence} in {Light} of {DESI} {DR2}},
	shorttitle = {Kinetic {Mixing} and the {Phantom} {Illusion}},
	url = {http://arxiv.org/abs/2511.23463},
	doi = {10.48550/arXiv.2511.23463},
	urldate = {2026-03-31},
	publisher = {arXiv},
	author = {Toomey, Michael W. and Hughes, Ellie and Ivanov, Mikhail M. and Sullivan, James M.},
	month = nov,
	year = {2025},
	note = {arXiv:2511.23463 [astro-ph]},
}

@misc{toomey_how_2025,
	title = {How {Theory}-{Informed} {Priors} {Affect} {DESI} {Evidence} for {Evolving} {Dark} {Energy}},
	url = {http://arxiv.org/abs/2509.13318},
	doi = {10.48550/arXiv.2509.13318},
	urldate = {2026-03-31},
	publisher = {arXiv},
	author = {Toomey, Michael W. and Montefalcone, Gabriele and McDonough, Evan and Freese, Katherine},
	month = sep,
	year = {2025},
	note = {arXiv:2509.13318 [astro-ph]},
}

@article{traykova_theoretical_2021,
	title = {Theoretical priors in scalar-tensor cosmologies: {Shift}-symmetric {Horndeski} models},
	volume = {104},
	shorttitle = {Theoretical priors in scalar-tensor cosmologies},
	doi = {10.1103/PhysRevD.104.083502},
	number = {8},
	journal = {Physical Review D},
	author = {Traykova, Dina},
	year = {2021},
}

@article{marsh_quintessence_2014,
	title = {Quintessence in a quandary: {Prior} dependence in dark energy models},
	volume = {90},
	shorttitle = {Quintessence in a quandary},
	doi = {10.1103/PhysRevD.90.105023},
	number = {10},
	journal = {Physical Review D},
	author = {Marsh, David J. E.},
	year = {2014},
}

@article{yang_probing_2025,
	title = {Probing the cold nature of dark matter},
	volume = {111},
	url = {https://link.aps.org/doi/10.1103/PhysRevD.111.103509},
	doi = {10.1103/PhysRevD.111.103509},
	number = {10},
	urldate = {2026-03-31},
	journal = {Physical Review D},
	author = {Yang, Weiqiang and Pan, Supriya and Di Valentino, Eleonora and Mena, Olga and Mota, David F. and Chakraborty, Subenoy},
	month = may,
	year = {2025},
	note = {Publisher: American Physical Society},
	pages = {103509},
}

@misc{kumar_evidence_2025,
	title = {Evidence for non-cold dark matter from {DESI} {DR2} measurements},
	url = {http://arxiv.org/abs/2504.14419},
	doi = {10.48550/arXiv.2504.14419},
	urldate = {2026-03-31},
	publisher = {arXiv},
	author = {Kumar, Utkarsh and Ajith, Abhijith and Verma, Amresh},
	month = apr,
	year = {2025},
	note = {arXiv:2504.14419 [astro-ph]},
}

@article{giani_matter_2025,
	title = {The matter with(in) {CPL}},
	volume = {8},
	issn = {2565-6120},
	url = {http://arxiv.org/abs/2505.08467},
	doi = {10.33232/001c.142699},
	urldate = {2026-03-31},
	journal = {The Open Journal of Astrophysics},
	author = {Giani, Leonardo and Marttens, Rodrigo Von and Piattella, Oliver Fabio},
	month = jul,
	year = {2025},
	note = {arXiv:2505.08467 [astro-ph]},
}

@misc{park_updated_2025,
	title = {Updated observational constraints on $\varphi${CDM} dynamical dark energy cosmological models},
	url = {http://arxiv.org/abs/2509.25812},
	doi = {10.48550/arXiv.2509.25812},
	urldate = {2026-03-31},
	publisher = {arXiv},
	author = {Park, Chan-Gyung and Ratra, Bharat},
	month = sep,
	year = {2025},
	note = {arXiv:2509.25812 [astro-ph]},
}

@article{giani_novel_2025,
	title = {Novel {Approach} to {Cosmological} {Nonlinearities} as an {Effective} {Fluid}},
	volume = {135},
	url = {https://link.aps.org/doi/10.1103/zr92-m7py},
	doi = {10.1103/zr92-m7py},
	number = {7},
	urldate = {2026-03-31},
	journal = {Physical Review Letters},
	author = {Giani, Leonardo and von Marttens, Rodrigo and Camilleri, Ryan},
	month = aug,
	year = {2025},
	note = {Publisher: American Physical Society},
	pages = {071004},
}

@misc{collaboration_local_2025,
	title = {The {Local} {Distance} {Network}: a community consensus report on the measurement of the {Hubble} constant at 1\% precision},
	shorttitle = {The {Local} {Distance} {Network}},
	url = {http://arxiv.org/abs/2510.23823},
	doi = {10.48550/arXiv.2510.23823},
	urldate = {2025-12-02},
	publisher = {arXiv},
	author = {Casertano, Stefano and others},
	month = oct,
	year = {2025},
	note = {arXiv:2510.23823 [astro-ph]},
	collaboration = {H0DN Collaboration},
}

@article{riess_comprehensive_2022,
	title = {A {Comprehensive} {Measurement} of the {Local} {Value} of the {Hubble} {Constant} with 1 km s−1 {Mpc}−1 {Uncertainty} from the {Hubble} {Space} {Telescope} and the {SH0ES} {Team}},
	volume = {934},
	issn = {2041-8205},
	url = {https://doi.org/10.3847/2041-8213/ac5c5b},
	doi = {10.3847/2041-8213/ac5c5b},
	language = {en},
	number = {1},
	urldate = {2025-12-02},
	journal = {The Astrophysical Journal Letters},
	author = {Riess, Adam G. and Yuan, Wenlong and Macri, Lucas M. and Scolnic, Dan and Brout, Dillon and Casertano, Stefano and Jones, David O. and Murakami, Yukei and Anand, Gagandeep S. and Breuval, Louise and Brink, Thomas G. and Filippenko, Alexei V. and Hoffmann, Samantha and Jha, Saurabh W. and D’arcy Kenworthy, W. and Mackenty, John and Stahl, Benjamin E. and Zheng, WeiKang},
	month = jul,
	year = {2022},
	zoteronote = {Publisher: The American Astronomical Society},
	pages = {L7},
}

@article{freedman_status_2025,
	title = {Status {Report} on the {Chicago}-{Carnegie} {Hubble} {Program} ({CCHP}): {Measurement} of the {Hubble} {Constant} {Using} the {Hubble} and {James} {Webb} {Space} {Telescopes}},
	volume = {985},
	issn = {0004-637X},
	shorttitle = {Status {Report} on the {Chicago}-{Carnegie} {Hubble} {Program} ({CCHP})},
	url = {https://doi.org/10.3847/1538-4357/adce78},
	doi = {10.3847/1538-4357/adce78},
	language = {en},
	number = {2},
	urldate = {2025-12-02},
	journal = {The Astrophysical Journal},
	author = {Freedman, Wendy L. and Madore, Barry F. and Hoyt, Taylor J. and Jang, In Sung and Lee, Abigail J. and Owens, Kayla A.},
	month = may,
	year = {2025},
	zoteronote = {Publisher: The American Astronomical Society},
	pages = {203},
}

@article{mehta_superradiance_2021,
	title = {Superradiance in string theory},
	volume = {2021},
	issn = {1475-7516},
	url = {https://doi.org/10.1088/1475-7516/2021/07/033},
	doi = {10.1088/1475-7516/2021/07/033},
	language = {en},
	number = {07},
	urldate = {2025-12-01},
	journal = {Journal of Cosmology and Astroparticle Physics},
	author = {Mehta, Viraf M. and Demirtas, Mehmet and Long, Cody and Marsh, David J.E. and McAllister, Liam and Stott, Matthew J.},
	month = jul,
	year = {2021},
	zoteronote = {Publisher: IOP Publishing},
	pages = {033},
}

@article{shlivko_tcc,
	title = {Trans-{Planckian} censorship constraints on properties and cosmological applications of axion-like fields},
	volume = {846},
	issn = {0370-2693},
	url = {https://www.sciencedirect.com/science/article/pii/S0370269323005853},
	doi = {10.1016/j.physletb.2023.138251},
	urldate = {2025-04-01},
	journal = {Physics Letters B},
	author = {Shlivko, David},
	month = nov,
	year = {2023},
	pages = {138251},
}

@misc{desi_forest_2025,
	title = {{DESI} {DR2} {Results} {I}: {Baryon} {Acoustic} {Oscillations} from the {Lyman} {Alpha} {Forest}},
	shorttitle = {{DESI} {DR2} {Results} {I}},
	url = {http://arxiv.org/abs/2503.14739},
	doi = {10.48550/arXiv.2503.14739},
	urldate = {2025-03-28},
	publisher = {arXiv},
	author = {Karim, M. A. and others},
	collaboration = {DESI Collaboration},
	month = mar,
	year = {2025},
	note = {arXiv:2503.14739 [astro-ph]},
	anzoteronote = {Comment: Submitted to PRD. Updated authors and references. 28 pages and 13 figures. This DESI Collaboration Publication is part of the Data Release 2 publication series (see https://data.desi.lbl.gov/doc/papers )},
}

@misc{desi_cosmo_2025,
	title = {{DESI} {DR2} {Results} {II}: {Measurements} of {Baryon} {Acoustic} {Oscillations} and {Cosmological} {Constraints}},
	shorttitle = {{DESI} {DR2} {Results} {II}},
	url = {http://arxiv.org/abs/2503.14738},
	doi = {10.48550/arXiv.2503.14738},
	urldate = {2025-03-21},
	publisher = {arXiv},
	author = {Karim, M. A. and others},
	collaboration = {DESI Collaboration},
	month = mar,
	year = {2025},
	note = {arXiv:2503.14738 [astro-ph]},
	anzoteronote = {Comment: 40 pages, 18 figures. This DESI Collaboration Publication is part of the Data Release 2 publication series (see https://data.desi.lbl.gov/doc/papers )},
}

@article{BIC_schwarz,
	title = {Estimating the {Dimension} of a {Model}},
	volume = {6},
	issn = {0090-5364, 2168-8966},
	url = {https://projecteuclid.org/journals/annals-of-statistics/volume-6/issue-2/Estimating-the-Dimension-of-a-Model/10.1214/aos/1176344136.full},
	doi = {10.1214/aos/1176344136},
	number = {2},
	urldate = {2024-12-18},
	journal = {The Annals of Statistics},
	author = {Schwarz, Gideon},
	month = mar,
	year = {1978},
	zoteronote = {Publisher: Institute of Mathematical Statistics},
	pages = {461--464},
}

@article{akaike_new_1974,
	title = {A new look at the statistical model identification},
	volume = {19},
	copyright = {https://ieeexplore.ieee.org/Xplorehelp/downloads/license-information/IEEE.html},
	issn = {0018-9286},
	url = {http://ieeexplore.ieee.org/document/1100705/},
	doi = {10.1109/TAC.1974.1100705},
	language = {en},
	number = {6},
	urldate = {2025-11-22},
	journal = {IEEE Transactions on Automatic Control},
	author = {Akaike, H.},
	month = dec,
	year = {1974},
	pages = {716--723},
}

@article{handley_quantifying_2019,
	title = {Quantifying dimensionality: {Bayesian} cosmological model complexities},
	volume = {100},
	shorttitle = {Quantifying dimensionality},
	url = {https://link.aps.org/doi/10.1103/PhysRevD.100.023512},
	doi = {10.1103/PhysRevD.100.023512},
	number = {2},
	urldate = {2025-11-22},
	journal = {Physical Review D},
	author = {Handley, Will and Lemos, Pablo},
	month = jul,
	year = {2019},
	zoteronote = {Publisher: American Physical Society},
	pages = {023512},
}

@article{liddle_information_2007,
	title = {Information criteria for astrophysical model selection},
	volume = {377},
	issn = {1745-3925},
	url = {https://doi.org/10.1111/j.1745-3933.2007.00306.x},
	doi = {10.1111/j.1745-3933.2007.00306.x},
	number = {1},
	urldate = {2024-12-18},
	journal = {Monthly Notices of the Royal Astronomical Society: Letters},
	author = {Liddle, Andrew R.},
	month = may,
	year = {2007},
	pages = {L74--L78},
}

@article{carron_Planck_lensing,
	title = {{CMB} lensing from {Planck} {PR4} maps},
	volume = {2022},
	issn = {1475-7516},
	url = {https://dx.doi.org/10.1088/1475-7516/2022/09/039},
	doi = {10.1088/1475-7516/2022/09/039},
	language = {en},
	number = {09},
	urldate = {2024-12-12},
	journal = {Journal of Cosmology and Astroparticle Physics},
	author = {Carron, Julien and Mirmelstein, Mark and Lewis, Antony},
	month = sep,
	year = {2022},
	zoteronote = {Publisher: IOP Publishing},
	pages = {039},
}

@article{ACT_lensing1,
	title = {The {Atacama} {Cosmology} {Telescope}: {A} {Measurement} of the {DR6} {CMB} {Lensing} {Power} {Spectrum} and {Its} {Implications} for {Structure} {Growth}},
	volume = {962},
	issn = {0004-637X},
	shorttitle = {The {Atacama} {Cosmology} {Telescope}},
	url = {https://dx.doi.org/10.3847/1538-4357/acfe06},
	doi = {10.3847/1538-4357/acfe06},
	language = {en},
	number = {2},
	urldate = {2024-12-12},
	journal = {The Astrophysical Journal},
	author = {Qu, Frank J. and others},
	collaboration = {ACT Collaboration},
	month = feb,
	year = {2024},
	zoteronote = {Publisher: The American Astronomical Society},
	pages = {112},
}

@article{ACT_lensing2,
	title = {The {Atacama} {Cosmology} {Telescope}: {Mitigating} the {Impact} of {Extragalactic} {Foregrounds} for the {DR6} {Cosmic} {Microwave} {Background} {Lensing} {Analysis}},
	volume = {966},
	issn = {0004-637X},
	shorttitle = {The {Atacama} {Cosmology} {Telescope}},
	url = {https://dx.doi.org/10.3847/1538-4357/ad2610},
	doi = {10.3847/1538-4357/ad2610},
	language = {en},
	number = {1},
	urldate = {2024-12-12},
	journal = {The Astrophysical Journal},
	author = {MacCrann, Niall and others},
	collaboration = {ACT Collaboration},
	month = apr,
	year = {2024},
	zoteronote = {Publisher: The American Astronomical Society},
	pages = {138},
}

@article{ACT_lensing3,
	title = {The {Atacama} {Cosmology} {Telescope}: {DR6} {Gravitational} {Lensing} {Map} and {Cosmological} {Parameters}},
	volume = {962},
	issn = {0004-637X},
	shorttitle = {The {Atacama} {Cosmology} {Telescope}},
	url = {https://dx.doi.org/10.3847/1538-4357/acff5f},
	doi = {10.3847/1538-4357/acff5f},
	language = {en},
	number = {2},
	urldate = {2024-12-12},
	journal = {The Astrophysical Journal},
	author = {Madhavacheril, Mathew S. and others},
	collaboration = {ACT Collaboration},
	month = feb,
	year = {2024},
	zoteronote = {Publisher: The American Astronomical Society},
	pages = {113},
}

@article{mead_hmcode_2015,
	title = {{HMcode}: {Halo}-model matter power spectrum computation},
	shorttitle = {{HMcode}},
	url = {https://ui.adsabs.harvard.edu/abs/2015ascl.soft08001M},
	urldate = {2024-12-11},
	journal = {Astrophysics Source Code Library},
	author = {Mead, Alexander},
	month = aug,
	year = {2015},
	zoteronote = {ADS Bibcode: 2015ascl.soft08001M},
	pages = {ascl:1508.001},
}

@article{mead_accurate_2016,
	title = {Accurate halo-model matter power spectra with dark energy, massive neutrinos and modified gravitational forces},
	volume = {459},
	issn = {0035-8711},
	url = {https://ui.adsabs.harvard.edu/abs/2016MNRAS.459.1468M},
	doi = {10.1093/mnras/stw681},
	urldate = {2024-12-11},
	journal = {Monthly Notices of the Royal Astronomical Society},
	author = {Mead, A. J. and Heymans, C. and Lombriser, L. and Peacock, J. A. and Steele, O. I. and Winther, H. A.},
	month = jun,
	year = {2016},
	zoteronote = {Publisher: OUP
ADS Bibcode: 2016MNRAS.459.1468M},
	pages = {1468--1488},
}

@article{fang_ppf,
	title = {Crossing the phantom divide with parametrized post-{Friedmann} dark energy},
	volume = {78},
	url = {https://link.aps.org/doi/10.1103/PhysRevD.78.087303},
	doi = {10.1103/PhysRevD.78.087303},
	number = {8},
	urldate = {2024-12-11},
	journal = {Physical Review D},
	author = {Fang, Wenjuan and Hu, Wayne and Lewis, Antony},
	month = oct,
	year = {2008},
	zoteronote = {Publisher: American Physical Society},
	pages = {087303},
}

@article{gelman_inference_1992,
	title = {Inference from {Iterative} {Simulation} {Using} {Multiple} {Sequences}},
	volume = {7},
	issn = {0883-4237, 2168-8745},
	url = {https://projecteuclid.org/journals/statistical-science/volume-7/issue-4/Inference-from-Iterative-Simulation-Using-Multiple-Sequences/10.1214/ss/1177011136.full},
	doi = {10.1214/ss/1177011136},
	number = {4},
	urldate = {2024-12-11},
	journal = {Statistical Science},
	author = {Gelman, Andrew and Rubin, Donald B.},
	month = nov,
	year = {1992},
	zoteronote = {Publisher: Institute of Mathematical Statistics},
	pages = {457--472},
}

@article{metropolis1,
	title = {Cosmological parameters from {CMB} and other data: {A} {Monte} {Carlo} approach},
	volume = {66},
	shorttitle = {Cosmological parameters from {CMB} and other data},
	url = {https://link.aps.org/doi/10.1103/PhysRevD.66.103511},
	doi = {10.1103/PhysRevD.66.103511},
	number = {10},
	urldate = {2024-12-11},
	journal = {Physical Review D},
	author = {Lewis, Antony and Bridle, Sarah},
	month = nov,
	year = {2002},
	zoteronote = {Publisher: American Physical Society},
	pages = {103511},
}

@article{metropolis2,
	title = {Efficient sampling of fast and slow cosmological parameters},
	volume = {87},
	url = {https://link.aps.org/doi/10.1103/PhysRevD.87.103529},
	doi = {10.1103/PhysRevD.87.103529},
	number = {10},
	urldate = {2024-12-11},
	journal = {Physical Review D},
	author = {Lewis, Antony},
	month = may,
	year = {2013},
	zoteronote = {Publisher: American Physical Society},
	pages = {103529},
}

@article{getdist,
	title = {{GetDist}: a {Python} package for analysing {Monte} {Carlo} samples},
	volume = {2025},
	issn = {1475-7516},
	shorttitle = {{GetDist}},
	url = {https://doi.org/10.1088/1475-7516/2025/08/025},
	doi = {10.1088/1475-7516/2025/08/025},
	language = {en},
	number = {08},
	urldate = {2026-04-04},
	journal = {Journal of Cosmology and Astroparticle Physics},
	author = {Lewis, Antony},
	month = aug,
	year = {2025},
	note = {Publisher: IOP Publishing},
	pages = {025},
}

@misc{metropolis_drag,
	title = {Taking {Bigger} {Metropolis} {Steps} by {Dragging} {Fast} {Variables}},
	url = {http://arxiv.org/abs/math/0502099},
	doi = {10.48550/arXiv.math/0502099},
	urldate = {2025-12-10},
	publisher = {arXiv},
	author = {Neal, Radford M.},
	month = feb,
	year = {2005},
	note = {arXiv:math/0502099},
}

@article{camb1,
      author         = "Lewis, Antony and Challinor, Anthony and Lasenby,
                        Anthony",
      title          = "{Efficient computation of CMB anisotropies in closed FRW
                        models}",
      journal        = "Astrophys. J.",
      volume         = "538",
      year           = "2000",
      pages          = "473-476",
      doi            = "10.1086/309179",
      eprint         = "astro-ph/9911177",
      archivePrefix  = "arXiv",
      primaryClass   = "astro-ph",
      SLACcitation   = "%%CITATION = ASTRO-PH/9911177;%%",
      url            = {https://arxiv.org/abs/astro-ph/9911177}
}

@article{camb2,
      author         = "Howlett, Cullan and Lewis, Antony and Hall, Alex and
                        Challinor, Anthony",
      title          = "{CMB power spectrum parameter degeneracies in the era of
                        precision cosmology}",
      journal        = "JCAP",
      volume         = "1204",
      year           = "2012",
      pages          = "027",
      doi            = "10.1088/1475-7516/2012/04/027",
      eprint         = "1201.3654",
      archivePrefix  = "arXiv",
      primaryClass   = "astro-ph.CO",
      SLACcitation   = "%%CITATION = ARXIV:1201.3654;%%",
      url            = {https://arxiv.org/abs/1201.3654}
}

@article{cobaya1,
    author = "Torrado, Jesus and Lewis, Antony",
    title = "{Cobaya: Code for Bayesian Analysis of hierarchical physical models}",
    eprint = "2005.05290",
    archivePrefix = "arXiv",
    primaryClass = "astro-ph.IM",
    reportNumber = "TTK-20-15",
    doi = "10.1088/1475-7516/2021/05/057",
    journal = "JCAP",
    volume = "05",
    pages = "057",
    year = "2021"
}

@article{cobaya2,
	title = {Cobaya: {Bayesian} analysis in cosmology},
	shorttitle = {Cobaya},
	url = {https://ui.adsabs.harvard.edu/abs/2019ascl.soft10019T},
	urldate = {2025-12-10},
	journal = {Astrophysics Source Code Library},
	author = {Torrado, Jesús and Lewis, Antony},
	month = oct,
	year = {2019},
	pages = {ascl:1910.019},
}

@misc{andrade_validation_2025,
	title = {Validation of the {DESI} {DR2} {Measurements} of {Baryon} {Acoustic} {Oscillations} from {Galaxies} and {Quasars}},
	url = {http://arxiv.org/abs/2503.14742},
	doi = {10.48550/arXiv.2503.14742},
	urldate = {2025-11-19},
	publisher = {arXiv},
	author = {Andrade, U. and others},
	collaboration = {DESI Collaboration},
	month = mar,
	year = {2025},
	note = {arXiv:2503.14742 [astro-ph]},
}

@misc{iminuit,
  author={Hans Dembinski and Piti Ongmongkolkul and others},
  title={scikit-hep/iminuit},
  DOI={10.5281/zenodo.3949207},
  publisher={Zenodo},
  year={2020},
  month={Dec},
  url={https://doi.org/10.5281/zenodo.3949207}
}

@article{handley_polychord_2015,
	title = {{PolyChord}: nested sampling for cosmology},
	volume = {450},
	issn = {1745-3933, 1745-3925},
	shorttitle = {{PolyChord}},
	url = {http://arxiv.org/abs/1502.01856},
	doi = {10.1093/mnrasl/slv047},
	number = {1},
	urldate = {2025-11-19},
	journal = {Monthly Notices of the Royal Astronomical Society: Letters},
	author = {Handley, W. J. and Hobson, M. P. and Lasenby, A. N.},
	month = jun,
	year = {2015},
	note = {arXiv:1502.01856 [astro-ph]},
	pages = {L61--L65},
}

@article{vincenzi_dark_2024,
	title = {The {Dark} {Energy} {Survey} {Supernova} {Program}: {Cosmological} {Analysis} and {Systematic} {Uncertainties}},
	volume = {975},
	issn = {0004-637X},
	shorttitle = {The {Dark} {Energy} {Survey} {Supernova} {Program}},
	url = {https://doi.org/10.3847/1538-4357/ad5e6c},
	doi = {10.3847/1538-4357/ad5e6c},
	language = {en},
	number = {1},
	urldate = {2025-11-21},
	journal = {The Astrophysical Journal},
	author = {Vincenzi, M. and others},
	collaboration = {DES Collaboration},
	month = oct,
	year = {2024},
	zoteronote = {Publisher: The American Astronomical Society},
	pages = {86},
}

@article{spiegelhalter_bayesian_2002,
	title = {Bayesian measures of model complexity and fit},
	volume = {64},
	doi = {10.1111/1467-9868.00353},
	number = {4},
	journal = {J. Roy. Statist. Soc. B},
	author = {Spiegelhalter, David J. and Best, Nicola G. and Carlin, Bradley P. and van der Linde, Angelika},
	year = {2002},
	pages = {583--639},
}

@article{spiegelhalter_deviance_2014,
	title = {The {Deviance} {Information} {Criterion}: 12 {Years} on},
	volume = {76},
	issn = {1369-7412},
	shorttitle = {The {Deviance} {Information} {Criterion}},
	url = {https://doi.org/10.1111/rssb.12062},
	doi = {10.1111/rssb.12062},
	number = {3},
	urldate = {2025-11-22},
	journal = {Journal of the Royal Statistical Society Series B: Statistical Methodology},
	author = {Spiegelhalter, David J. and Best, Nicola G. and Carlin, Bradley P. and Linde, Angelika},
	month = jun,
	year = {2014},
	pages = {485--493},
}

@article{sanchez_dark_2024,
	title = {The {Dark} {Energy} {Survey} {Supernova} {Program}: {Light} {Curves} and 5 {Yr} {Data} {Release}},
	volume = {975},
	issn = {0004-637X},
	shorttitle = {The {Dark} {Energy} {Survey} {Supernova} {Program}},
	url = {https://doi.org/10.3847/1538-4357/ad739a},
	doi = {10.3847/1538-4357/ad739a},
	language = {en},
	number = {1},
	urldate = {2025-11-21},
	journal = {The Astrophysical Journal},
	author = {S\'anchez, B. O. and others},
	collaboration = {DES Collaboration},
	month = oct,
	year = {2024},
	zoteronote = {Publisher: The American Astronomical Society},
	pages = {5},
}

@misc{naess_atacama_2025,
	title = {The {Atacama} {Cosmology} {Telescope}: {DR6} {Maps}},
	shorttitle = {The {Atacama} {Cosmology} {Telescope}},
	url = {http://arxiv.org/abs/2503.14451},
	doi = {10.48550/arXiv.2503.14451},
	urldate = {2025-11-21},
	publisher = {arXiv},
	author = {Naess, S. and others},
	collaboration = {ACT Collaboration},
	month = mar,
	year = {2025},
	note = {arXiv:2503.14451 [astro-ph]},
}

@article{handley_polychord_2015-1,
	title = {{PolyChord}: next-generation nested sampling},
	volume = {453},
	issn = {0035-8711, 1365-2966},
	shorttitle = {{PolyChord}},
	url = {http://arxiv.org/abs/1506.00171},
	doi = {10.1093/mnras/stv1911},
	number = {4},
	urldate = {2025-11-19},
	journal = {Monthly Notices of the Royal Astronomical Society},
	author = {Handley, W. J. and Hobson, M. P. and Lasenby, A. N.},
	month = nov,
	year = {2015},
	note = {arXiv:1506.00171 [astro-ph]},
	pages = {4385--4399},
}

@misc{popovic_dark_2025,
	title = {The {Dark} {Energy} {Survey} {Supernova} {Program}: {A} {Reanalysis} {Of} {Cosmology} {Results} {And} {Evidence} {For} {Evolving} {Dark} {Energy} {With} {An} {Updated} {Type} {Ia} {Supernova} {Calibration}},
	shorttitle = {The {Dark} {Energy} {Survey} {Supernova} {Program}},
	url = {http://arxiv.org/abs/2511.07517},
	doi = {10.48550/arXiv.2511.07517},
	urldate = {2025-11-19},
	publisher = {arXiv},
	author = {Popovic, B. and others},
	collaboration = {DES Collaboration},
	month = nov,
	year = {2025},
	note = {arXiv:2511.07517 [astro-ph]},
}

@article{casas_validation_2025,
	title = {Validation of the {DESI} {DR2} {Ly}$\alpha$ {BAO} analysis using synthetic datasets},
	volume = {113},
	url = {https://link.aps.org/doi/10.1103/fvgh-kswf},
	doi = {10.1103/fvgh-kswf},
	number = {2},
	urldate = {2026-04-04},
	journal = {Physical Review D},
	author = {Casas, L. and others},
	collaboration = {DESI Collaboration},
	month = jan,
	year = {2026},
	note = {Publisher: American Physical Society},
	pages = {023520},
}

@misc{brodzeller_construction_2025,
	title = {Construction of the {Damped} {Ly}$\alpha$ {Absorber} {Catalog} for {DESI} {DR2} {Ly}$\alpha$ {BAO}},
	url = {http://arxiv.org/abs/2503.14740},
	doi = {10.48550/arXiv.2503.14740},
	urldate = {2025-11-19},
	publisher = {arXiv},
	author = {Brodzeller, A. and others},
	collaboration = {DESI Collaboration},
	month = jun,
	year = {2025},
	note = {arXiv:2503.14740 [astro-ph]},
}

@article{pagano_reionization_2020,
	title = {Reionization optical depth determination from {Planck} {HFI} data with ten percent accuracy},
	volume = {635},
	copyright = {© ESO 2020},
	issn = {0004-6361, 1432-0746},
	url = {https://www.aanda.org/articles/aa/abs/2020/03/aa36630-19/aa36630-19.html},
	doi = {10.1051/0004-6361/201936630},
	language = {en},
	urldate = {2025-11-19},
	journal = {Astronomy \& Astrophysics},
	author = {Pagano, L. and Delouis, J.-M. and Mottet, S. and Puget, J.-L. and Vibert, L.},
	month = mar,
	year = {2020},
	zoteronote = {Publisher: EDP Sciences},
	pages = {A99},
}

@misc{louis_atacama_2025,
	title = {The {Atacama} {Cosmology} {Telescope}: {DR6} {Power} {Spectra}, {Likelihoods} and $\Lambda${CDM} {Parameters}},
	shorttitle = {The {Atacama} {Cosmology} {Telescope}},
	url = {http://arxiv.org/abs/2503.14452},
	doi = {10.48550/arXiv.2503.14452},
	urldate = {2025-11-19},
	publisher = {arXiv},
	author = {Louis, T. and others},
	collaboration = {ACT Collaboration},
	month = jun,
	year = {2025},
	note = {arXiv:2503.14452 [astro-ph]},
}

@article{planck_2020_like,
	title = {Planck 2018 results - {V}. {CMB} power spectra and likelihoods},
	volume = {641},
	copyright = {© Planck Collaboration 2020},
	issn = {0004-6361, 1432-0746},
	url = {https://www.aanda.org/articles/aa/abs/2020/09/aa36386-19/aa36386-19.html},
	doi = {10.1051/0004-6361/201936386},
	language = {en},
	urldate = {2024-12-11},
	journal = {Astronomy \& Astrophysics},
	author = {Aghanim, N. and others},
	collaboration = {Planck Collaboration},
	month = sep,
	year = {2020},
	zoteronote = {Publisher: EDP Sciences},
	pages = {A5},
}

@article{nojiri_phantom_2025,
	title = {Phantom crossing and oscillating dark energy with \${F}({R})\$ gravity},
	volume = {112},
	url = {https://link.aps.org/doi/10.1103/16yg-966k},
	doi = {10.1103/16yg-966k},
	number = {10},
	urldate = {2026-04-01},
	journal = {Physical Review D},
	author = {Nojiri, S. and Odintsov, S. D. and Oikonomou, V. K.},
	month = nov,
	year = {2025},
	note = {Publisher: American Physical Society},
	pages = {104035},
}

@misc{koutroulis_uv-complete_2026,
	title = {{UV}-complete and stable {Quintom} {Dark} {Energy} models in the light of {DESI} {DR2}},
	url = {http://arxiv.org/abs/2603.24685},
	doi = {10.48550/arXiv.2603.24685},
	urldate = {2026-04-03},
	publisher = {arXiv},
	author = {Koutroulis, Fotis},
	month = mar,
	year = {2026},
	note = {arXiv:2603.24685 [hep-ph]},
}

@misc{nojiri_apparent_2026,
	title = {Apparent {Phantom} {Crossing} in {Gauss}-{Bonnet} {Gravity}},
	url = {http://arxiv.org/abs/2512.06279},
	doi = {10.48550/arXiv.2512.06279},
	urldate = {2026-04-01},
	publisher = {arXiv},
	author = {Nojiri, Shin'ichi and Odintsov, Sergei D. and Oikonomou, V. K.},
	month = mar,
	year = {2026},
	note = {arXiv:2512.06279 [gr-qc]},
}

@misc{lopez_non-minimally_2025,
	title = {Non-{Minimally} {Coupled} {Quintessence} in {Light} of {DESI}},
	url = {http://arxiv.org/abs/2510.14941},
	doi = {10.48550/arXiv.2510.14941},
	urldate = {2026-01-06},
	publisher = {arXiv},
	author = {L\'opez, Samuel Sánchez and Karam, Alexandros and Hazra, Dhiraj Kumar},
	month = oct,
	year = {2025},
	note = {arXiv:2510.14941 [astro-ph]},
}

@article{banerjee_hubble_2021,
	title = {Hubble sinks in the low-redshift swampland},
	volume = {103},
	url = {https://link.aps.org/doi/10.1103/PhysRevD.103.L081305},
	doi = {10.1103/PhysRevD.103.L081305},
	number = {8},
	urldate = {2026-01-06},
	journal = {Physical Review D},
	author = {Banerjee, A. and Cai, H. and Heisenberg, L. and Colgáin, E. O and Sheikh-Jabbari, M. M. and Yang, T.},
	month = apr,
	year = {2021},
	note = {Publisher: American Physical Society},
	pages = {L081305},
}

@article{payeur_observations_2025,
	title = {Do observations prefer thawing quintessence?},
	volume = {111},
	url = {https://link.aps.org/doi/10.1103/bggr-61nr},
	doi = {10.1103/bggr-61nr},
	number = {12},
	urldate = {2026-01-06},
	journal = {Physical Review D},
	author = {Payeur, Guillaume and McDonough, Evan and Brandenberger, Robert},
	month = jun,
	year = {2025},
	note = {Publisher: American Physical Society},
	pages = {123541},
}

@article{wolf_scant_2024,
	title = {Scant evidence for thawing quintessence},
	volume = {110},
	url = {https://link.aps.org/doi/10.1103/PhysRevD.110.083528},
	doi = {10.1103/PhysRevD.110.083528},
	number = {8},
	urldate = {2025-05-14},
	journal = {Physical Review D},
	author = {Wolf, William J. and García-García, Carlos and Bartlett, Deaglan J. and Ferreira, Pedro G.},
	month = oct,
	year = {2024},
	zoteronote = {Publisher: American Physical Society},
	pages = {083528},
}

@article{andriot_phantom_2025,
	title = {Phantom matters},
	volume = {49},
	issn = {2212-6864},
	url = {https://www.sciencedirect.com/science/article/pii/S2212686425001931},
	doi = {10.1016/j.dark.2025.102000},
	urldate = {2025-07-24},
	journal = {Physics of the Dark Universe},
	author = {Andriot, David},
	month = sep,
	year = {2025},
	pages = {102000},
}

@misc{khoury_apparent_2025,
	title = {Apparent \$w{\textless}-1\$ and a {Lower} \${S}\_8\$ from {Dark} {Axion} and {Dark} {Baryons} {Interactions}},
	url = {http://arxiv.org/abs/2503.16415},
	doi = {10.48550/arXiv.2503.16415},
	urldate = {2025-09-20},
	publisher = {arXiv},
	author = {Khoury, Justin and Lin, Meng-Xiang and Trodden, Mark},
	month = mar,
	year = {2025},
	note = {arXiv:2503.16415 [astro-ph]},
	anzoteronote = {Comment: 8 pages, 4 figures},
}

@article{PADE_alho,
	title = {New simple and accurate quintessence approximations},
	volume = {111},
	url = {https://link.aps.org/doi/10.1103/PhysRevD.111.083549},
	doi = {10.1103/PhysRevD.111.083549},
	number = {8},
	urldate = {2025-05-14},
	journal = {Physical Review D},
	author = {Alho, Artur and Uggla, Claes},
	month = apr,
	year = {2025},
	zoteronote = {Publisher: American Physical Society},
	pages = {083549},
}

@article{cahn_field_2008,
	title = {Field {Flows} of {Dark} {Energy}},
	volume = {11},
	doi = {10.1088/1475-7516/2008/11/015},
	journal = {JCAP},
	author = {Cahn, Robert N. and de Putter, Roland and Linder, Eric V.},
	year = {2008},
	zoteronote = {\_eprint: 0807.1346},
	pages = {015},
}

@article{assessing,
	title = {Assessing observational constraints on dark energy},
	volume = {855},
	copyright = {All rights reserved},
	issn = {0370-2693},
	url = {https://www.sciencedirect.com/science/article/pii/S0370269324003848},
	doi = {10.1016/j.physletb.2024.138826},
	urldate = {2024-07-16},
	journal = {Physics Letters B},
	author = {Shlivko, David and Steinhardt, Paul J.},
	month = aug,
	year = {2024},
	pages = {138826},
}

@article{wolf_underdetermination_2023,
    author = "Wolf, William J. and Ferreira, Pedro G.",
    title = "{Underdetermination of dark energy}",
    eprint = "2310.07482",
    archivePrefix = "arXiv",
    primaryClass = "astro-ph.CO",
    realdoi = "10.1103/PhysRevD.108.103519",
    journal = "Phys. Rev. D",
    volume = "108",
    number = "10",
    pages = "103519",
    year = "2023"
}

@article{garcia-garcia_theoretical_2020,
	title = {Theoretical priors in scalar-tensor cosmologies: {Thawing} quintessence},
	volume = {101},
	issn = {2470-0010, 2470-0029},
	shorttitle = {Theoretical priors in scalar-tensor cosmologies},
	realdoi = {10.1103/PhysRevD.101.063508},
	language = {en},
	number = {6},
	urldate = {2024-05-10},
	journal = {Physical Review D},
	author = {García-García, Carlos and Bellini, Emilio and Ferreira, Pedro G. and Traykova, Dina and Zumalacárregui, Miguel},
	realmonth = mar,
	year = {2020},
	pages = {063508},
}

@misc{bedroya_evolving_2025,
	title = {Evolving {Dark} {Sector} and the {Dark} {Dimension} {Scenario}},
	url = {http://arxiv.org/abs/2507.03090},
	doi = {10.48550/arXiv.2507.03090},
	urldate = {2025-08-15},
	publisher = {arXiv},
	author = {Bedroya, Alek and Obied, Georges and Vafa, Cumrun and Wu, David H.},
	month = jul,
	year = {2025},
	note = {arXiv:2507.03090 [astro-ph]},
	anzoteronote = {Comment: 13 pages, 7 figures},
}

@article{shlivko_optimal_2025,
		title = {Optimal parameterizations for observational constraints on thawing dark energy},
		volume = {2025},
		copyright = {All rights reserved},
		issn = {1475-7516},
		url = {https://dx.doi.org/10.1088/1475-7516/2025/06/054},
		doi = {10.1088/1475-7516/2025/06/054},
		language = {en},
		number = {06},
		urldate = {2025-06-25},
		journal = {Journal of Cosmology and Astroparticle Physics},
		author = {Shlivko, David and Steinhardt, Paul J. and Steinhardt, Charles L.},
		month = jun,
		year = {2025},
		zoteronote = {Publisher: IOP Publishing},
		pages = {054},
	}

@article{Ooguri:2018wrx,
    author = "Ooguri, Hirosi and Palti, Eran and Shiu, Gary and Vafa, Cumrun",
    title = "{Distance and de Sitter Conjectures on the Swampland}",
    eprint = "1810.05506",
    archivePrefix = "arXiv",
    primaryClass = "hep-th",
    realdoi = "10.1016/j.physletb.2018.11.018",
    journal = "Phys. Lett. B",
    volume = "788",
    pages = "180--184",
    year = "2019"
}

@article{DESI:2024vi,
	title = {{DESI} 2024 {VI}: cosmological constraints from the measurements of baryon acoustic oscillations},
	volume = {2025},
	issn = {1475-7516},
	shorttitle = {{DESI} 2024 {VI}},
	url = {https://dx.doi.org/10.1088/1475-7516/2025/02/021},
	doi = {10.1088/1475-7516/2025/02/021},
	language = {en},
	number = {02},
	urldate = {2025-05-14},
	journal = {Journal of Cosmology and Astroparticle Physics},
	author = {Adame, A.G. and others},
	collaboration = {DESI Collaboration},
	month = feb,
	year = {2025},
	pages = {021},
}

@article{SN:pantheon,
	title = {The {Pantheon}+ {Analysis}: {Cosmological} {Constraints}},
	volume = {938},
	issn = {0004-637X},
	shorttitle = {The {Pantheon}+ {Analysis}},
	url = {https://dx.doi.org/10.3847/1538-4357/ac8e04},
	doi = {10.3847/1538-4357/ac8e04},
	language = {en},
	number = {2},
	urldate = {2024-12-11},
	journal = {The Astrophysical Journal},
	author = {Brout, D. and others},
	month = oct,
	year = {2022},
	pages = {110},
}

@article{SN:union,
	title = {Union through {UNITY}: {Cosmology} with 2000 {SNe} {Using} a {Unified} {Bayesian} {Framework}},
	volume = {986},
	issn = {0004-637X},
	shorttitle = {Union through {UNITY}},
	url = {https://doi.org/10.3847/1538-4357/adc0a5},
	doi = {10.3847/1538-4357/adc0a5},
	language = {en},
	number = {2},
	urldate = {2025-12-10},
	journal = {The Astrophysical Journal},
	author = {Rubin, David and Aldering, Greg and Betoule, Marc and Fruchter, Andy and Huang, Xiaosheng and Kim, Alex G. and Lidman, Chris and Linder, Eric and Perlmutter, Saul and Ruiz-Lapuente, Pilar and Suzuki, Nao},
	month = jun,
	year = {2025},
	pages = {231},
}

@article{SN:DES,
	title = {The {Dark} {Energy} {Survey}: {Cosmology} {Results} with \textasciitilde{}1500 {New} {High}-redshift {Type} {Ia} {Supernovae} {Using} the {Full} 5 yr {Data} {Set}},
	volume = {973},
	issn = {2041-8205},
	shorttitle = {The {Dark} {Energy} {Survey}},
	url = {https://dx.doi.org/10.3847/2041-8213/ad6f9f},
	doi = {10.3847/2041-8213/ad6f9f},
	language = {en},
	number = {1},
	urldate = {2025-05-14},
	journal = {The Astrophysical Journal Letters},
	author = {Abbott, T. M. C. and others},
	collaboration = {DES Collaboration},
	month = oct,
	year = {2024},
	zoteronote = {Publisher: The American Astronomical Society},
	pages = {L14},
}

@article{SN:DES_collab,
	title = {The {Dark} {Energy} {Survey}: {Cosmology} {Results} with \textasciitilde{}1500 {New} {High}-redshift {Type} {Ia} {Supernovae} {Using} the {Full} 5 yr {Data} {Set}},
	volume = {973},
	issn = {2041-8205},
	shorttitle = {The {Dark} {Energy} {Survey}},
	url = {https://dx.doi.org/10.3847/2041-8213/ad6f9f},
	doi = {10.3847/2041-8213/ad6f9f},
	language = {en},
	number = {1},
	urldate = {2025-05-14},
	journal = {The Astrophysical Journal Letters},
	author = {Abbott, T. M. C. and others},
	collaboration = {DES Collaboration},
	month = oct,
	year = {2024},
	zoteronote = {Publisher: The American Astronomical Society},
	pages = {L14},
}

@misc{adam_comparing_2025,
	title = {Comparing {Minimal} and {Non}-{Minimal} {Quintessence} {Models} to 2025 {DESI} {Data}},
	url = {http://arxiv.org/abs/2509.13302},
	doi = {10.48550/arXiv.2509.13302},
	urldate = {2025-09-20},
	publisher = {arXiv},
	author = {Adam, Husam and Hertzberg, Mark P. and Jiménez-Aguilar, Daniel and Khan, Iman},
	month = sep,
	year = {2025},
	note = {arXiv:2509.13302 [astro-ph]},
	anzoteronote = {Comment: 29 pages, 9 figures, 11 tables},
}

@misc{chen_quintessential_2025,
	title = {Quintessential dark energy crossing the phantom divide},
	url = {http://arxiv.org/abs/2508.19101},
	doi = {10.48550/arXiv.2508.19101},
	urldate = {2025-09-20},
	publisher = {arXiv},
	author = {Chen, Ruiqi and Cline, James M. and Muralidharan, Varun and Salewicz, Benjamin},
	month = aug,
	year = {2025},
	note = {arXiv:2508.19101 [astro-ph]},
	anzoteronote = {Comment: 10 pages, 6 figures},
}

@article{de_putter_calibrating_2008,
	title = {Calibrating dark energy},
	volume = {2008},
	issn = {1475-7516},
	url = {https://doi.org/10.1088/1475-7516/2008/10/042},
	doi = {10.1088/1475-7516/2008/10/042},
	language = {en},
	number = {10},
	urldate = {2025-12-18},
	journal = {Journal of Cosmology and Astroparticle Physics},
	author = {de Putter, Roland and Linder, Eric V},
	month = oct,
	year = {2008},
	pages = {042},
}

@article{chen_evolving_2025,
	title = {Evolving dark energy or dark matter with an evolving equation-of-state?},
	volume = {2025},
	issn = {1475-7516},
	url = {https://dx.doi.org/10.1088/1475-7516/2025/07/059},
	doi = {10.1088/1475-7516/2025/07/059},
	language = {en},
	number = {07},
	urldate = {2025-07-24},
	journal = {Journal of Cosmology and Astroparticle Physics},
	author = {Chen, Xingang and Loeb, Abraham},
	month = jul,
	year = {2025},
	zoteronote = {Publisher: IOP Publishing},
	pages = {059},
}

@misc{brax_weinbergs_2025,
	title = {Weinberg's theorem, phantom crossing and screening},
	url = {http://arxiv.org/abs/2507.16723},
	doi = {10.48550/arXiv.2507.16723},
	urldate = {2025-07-25},
	publisher = {arXiv},
	author = {Brax, Philippe},
	month = jul,
	year = {2025},
	note = {arXiv:2507.16723 [astro-ph]},
	anzoteronote = {Comment: 14 pages, 1 figure},
}

@article{wolf_assessing_2025,
	title = {Assessing {Cosmological} {Evidence} for {Nonminimal} {Coupling}},
	volume = {135},
	url = {https://link.aps.org/doi/10.1103/jysf-k72m},
	doi = {10.1103/jysf-k72m},
	number = {8},
	urldate = {2025-12-09},
	journal = {Physical Review Letters},
	author = {Wolf, William J. and García-García, Carlos and Anton, Theodore and Ferreira, Pedro G.},
	month = aug,
	year = {2025},
	zoteronote = {Publisher: American Physical Society},
	pages = {081001},
}

\end{document}